\documentclass[sn-basic]{sn-jnl}

\usepackage{graphicx}%
\usepackage{multirow}%
\usepackage{amsmath,amssymb,amsfonts}%
\usepackage{amsthm}%
\usepackage{mathrsfs}%
\usepackage[title]{appendix}%
\usepackage{xcolor}%
\usepackage{textcomp}%
\usepackage{manyfoot}%
\usepackage{booktabs}%
\usepackage{algorithm}%
\usepackage{algorithmicx}%
\usepackage{algpseudocode}%
\usepackage{listings}%
\usepackage{array}

\newcommand{\permil}{\mbox{\textperthousand}}
\newcommand{\gtsimeq}{\raisebox{-0.6ex}{$\, \stackrel{\raisebox{-.2ex}
{$\textstyle >$}}{\sim}\,$}}
\newcommand{\ltsimeq}{\raisebox{-0.6ex}{$\, \stackrel{\raisebox{-.2ex}
{$\textstyle <$}}{\sim}\,$}}

\newenvironment{Steve}{\color{black}}{\ignorespacesafterend}

\begin{document}

\title[Article Title]{The Sun's Birth Environment:\linebreak Context for Meteoritics}


\author*[1]{\fnm{Steve} \sur{Desch}}\email{steve.desch@asu.edu}

\author[2]{\fnm{N\'{u}ria} \sur{Miret-Roig}}\email{nuria.miret.roig@univie.ac.at}
%

\affil*[1]{\orgdiv{School of Earth and Space Exploration}, \orgname{Arizona State University}, \linebreak \orgaddress{\street{PO Box 876004}, \city{Tempe}, \state{Arizona} \postcode{85287-6004}, \country{USA}}}

\affil[2]{\orgdiv{Department of Astrophysics}, \orgname{University of Vienna}, \linebreak \orgaddress{\street{T\"{u}rkenschanzstrasse 17 (Sternwarte)},  \city{1180 Wien}, \country{Austria}}}



\abstract{
Meteorites trace planet formation in the Sun's protoplanetary disk, but they also record the influence of the Sun's birth environment. 
Whether the Sun formed in a region like Taurus-Auriga with \mbox{$\sim 10^2$} stars, or a region like the Carina Nebula with \mbox{$\sim 10^6$} stars, matters for how large the Sun's disk was, for how long and from how far away it accreted gas from the molecular cloud, and how it acquired radionuclides like \mbox{${}^{26}{\rm Al}$}. 
To provide context for the interpretation of meteoritic data, we review what is known about the Sun's birth environment. 
Based on an inferred gas disk outer radius $\approx 50 - 90$ AU, radial transport in the disk, and the abundances of noble gases in Jupiter's atmosphere, the Sun's molecular cloud and protoplanetary disk was exposed to an ultraviolet flux $G_0 \sim 30 - 3000$ during its birth and first $\approx 10$ Myr of evolution.
Based on the orbits of Kuiper Belt objects, the Solar System was subsequently exposed to a stellar density $\approx \, 100 \, M_{\odot} \, {\rm pc}^{-3}$ for $\approx 100$ Myr, strongly implying formation in a bound cluster. 
These facts suggest formation in a region like the outskirts of the Orion Nebula, perhaps 2 pc from the center.
The protoplanetary disk might have accreted gas for many Myr, but a few \mbox{$\times 10^5$} yr seems more likely. 
It probably inherited radionuclides from its molecular cloud, enriched by inputs from supernovae and especially Wolf-Rayet star winds, and acquired a typical amount of \mbox{${}^{26}{\rm Al}$}. 
}




\maketitle

\section{Introduction}\label{sec1}

Today the Sun and the planets in their orbits evolve slowly, due to the
internal workings of the Solar System, isolated from the goings-on in the Galaxy.
Not so at its birth, when the effects of the early Sun's surroundings
determined how the Solar System would evolve for the next billions of years.
All new solar systems are greatly affected by their birth environments,
especially the number of stars, $N$, in their birth clusters, which can range
from $N \sim 10^2$ stars in low-mass star-forming regions like Taurus-Auriga
and Lupus, to $\sim 10^3$ stars in high-mass star-forming regions like
Upper Scorpius (Upper Sco), to $\sim 10^4$ stars in regions like the
Orion Nebula Cluster (ONC), to $\sim 10^6$ stars in regions like the Carina Nebula.
A major goal of Solar System studies is to determine the number of stars in
the Sun's birth environment, and the effects those stars had on its future
evolution.

About as many stars form in regions with $N \sim 10^2$ as in clusters with
$N \sim 10^4$ or $N \sim 10^6$, but their formation and evolution will
proceed very differently, because of the different stellar densities and
especially ultraviolet (UV) fluxes \citep{LadaLada2003,HesterDesch2005,Adams2010}.
Clusters with large $N$ have a greater likelihood of randomly generating
high-mass O and B stars, which will emit tremendous amounts of UV that can
ionize and heat the gas, creating a so-called H {\sc ii} region.
A single O6 star like $\theta^1$ Ori C in the ONC emits $5 \times 10^5$ times the
luminosity of the Sun, half of that energy in the UV.
The ratio of the UV flux to the background UV flux in the Galaxy ($1.6 \times 10^{-3} \, {\rm erg} \, {\rm cm}^{-2} \, {\rm s}^{-1}$) is denoted $G_0$.
Regions like Taurus-Auriga have $N \sim 10^2$, lack OB stars and are characterized
by $G_0 \sim 1$ and stellar densities $\rho_{\star} \sim 1 \, M_{\odot} \, {\rm pc}^{-3}$
\citep{GudelEtal2007}.
Regions like Upper Sco have $N \sim 10^3$, dozens of B stars and $G_0 \sim 10^3$,
and $\rho_{\star} \sim 1 \, M_{\odot} \, {\rm pc}^{-3}$ \citep{Luhman2020a}.
The adjacent $\rho$ Ophiuchi ($\rho$ Oph) region has $N \sim 10^3$ stars,
including 2 B stars, $G_0 = 57^{+202}_{-31}$ and
 $\rho_{\star} \sim 13^{+25}_{-6} \, M_{\odot} \, {\rm pc}^{-3}$
\citep{ParkerEtal2012}.
The ONC, with $\theta^1$ Ori C, a few O stars, and a dozen B stars, sees $G_0 \sim 5000 \, (r / 1 \, {\rm pc})^{-2}$, where $r$ is the distance from the cluster center ($\theta^1$ Ori C)
\citep{StorzerHollenbach1999}.
The stellar density is $3.3 \times 10^4 \, M_{\odot} \, {\rm pc}^{-3}$ at the cluster center, and the average density within 2.06 pc is $49 \, M_{\odot} \, {\rm pc}^{-3}$ \citep{HillenbrandHartmann1998}.
%
And the most extreme locales like the Trumpler 14 subregion of the $N \sim 10^6$
Carina Nebula by themselves have $N \sim 10^3 - 10^4$, dozens of O stars,
$G_0 \sim 10^6$, and $\rho_{\star} > 10^3 \, M_{\odot} \, {\rm pc}^{-3}$ \citep{MesaDelgadoEtal2016}.
A priori, the Sun is equally likely to form in any of these environments \citep{Adams2010}.

The outcomes of formation in these different environments are quite divergent, starting
with the sizes of protoplanetary disks.
As reviewed by \citet{MarchingtonParker2022}, these are greatly affected by external
photoevaporation by the UV flux of nearby massive stars.
In low-mass star-forming regions like the $\sim 6$ Myr-old Lupus cloud, the median
radius of dust emission is about 100 AU.
In the similarly old (3--10 Myr; \cite{RatzenbockEtal2023, MiretRoigEtal2022}) Upper Sco region, 
the median dust radius is 40 AU;
but it is only 15 AU for disks (`proplyds') in the 2.5 Myr-old ONC.
In the extreme environment of Trumpler 14, no disks $> 3$ AU in radius are observed,
despite its young age (few $\times 10^5$ yr); planet formation is effectively shut
off in these systems \citep{MesaDelgadoEtal2016}.

The same UV flux photoevaporates the molecular gas that would feed protostars and
their disks.
In low-mass star-forming regions, even older Class II (few Myr old) protostars
and their disks remain embedded in their natal molecular clouds and may continue
to be fed by streamers of gas, as are AB Aur, GM Aur, SU Aur, and DG Tau
\citep{PinedaEtal2023}.
The proplyds in the ONC or Trumpler 14 H {\sc ii} regions are completely isolated
from their molecular clouds in $< 10^5$ yr \citep{MesaDelgadoEtal2016}.

It is not clear how quickly accretion is shut off to disks in regions of intermediate
mass like Upper Sco or in the Orion Nebula far from the cluster center.
In all H {\sc ii} regions, massive stars launch ionization fronts that propagate outward
leaving in their wake ionized and heated gas that then leaves the cluster.
A very massive O star can clear out an entire parsecs-wide cluster in as little as $\sim 10^5$ yr after its formation \citep{WallEtal2020}.
{What follows after that is a D-type ionization front in which a shock front propagates into the molecular cloud at speeds of a few km/s (few pc/Myr), compressing it; several tenths of a parsec behind the shock front, an ionization eats into the compressed gas, heating and ionizing it.}
The duration of accretion onto a protostar depends on the relative timing between its formation and when these ionization fronts pass it.
If the shock preceding a D-type ionization front {triggered} the formation of the
protostar, the relative timing is short, $\sim 10^5$ yr \citep{HesterDesch2005}.
Theoretical models \citep{DaleEtal2007,WallEtal2020} and observations 
\citep{SniderEtal2009,KarrMartin2003} suggest a substantial fraction ($\approx 25\%$)
of stars in H {\sc ii} regions were triggered to form in this way.
But the more-abundant, spontaneously-formed stars several pc from the cluster center may have to wait
many Myr to be uncovered.


Models of planet formation and the interpretation of meteoritic data require
pinning down which of these birth environments the Sun formed in.
For example, it has been suggested that planet-forming pebbles drifted in from $> 100$ AU
\citep{LambrechtsJohansen2014}, which is only possible in large disks.
Likewise, it has been suggested that the isotopic dichotomy in the Solar System
\citep{TrinquierEtal2009,Warren2011,KruijerEtal2017} is due to late infall
of material from the molecular cloud \citep{LiuEtal2022},
possibly along streamers \citep{ArzoumanianEtal2023}, which would require small $G_0$.
The Solar System started with live ${}^{26}{\rm Al}$ that was the cause of
asteroidal melting \citep{GrimmMcSween1993,LichtenbergEtal2019}.
It is commonly 
{(but not universally, see e.g., 
\citealt{JuraEtal2013,Young2014,DeschEtal2023c})
}
assumed that solar systems born with sufficient ${}^{26}{\rm Al}$
to melt asteroids might be rare \citep{CieslaEtal2015,LichtenbergEtal2019}
unless they formed in an especially large birth cluster \citep{WilliamsGaidos2007}.
Until recently, our intuition of how stars form was informed by studies of
low-mass star-forming regions like Taurus-Auriga, because this is the closest (140 pc),
most easily-observed star-forming region (cf. \citealt{HesterDesch2005}), and these
starting assumptions would be reasonable for such an environment.
But it has been increasingly recognized that stars like the Sun have a high
probability of forming in high-mass star-forming regions \citep{LadaLada2003,Adams2010},
and the signatures of such an origin have been sought in the Solar System.

To provide the proper context for disk and planet formation models and interpretation 
of meteoritic data, we are compelled to ask: 
\begin{itemize}
\item How large was the Sun's protoplanetary disk? 
\item How long did infall from the molecular cloud last? 
\item Was all this material local to the Sun's cloud core (and thus homogeneous), or brought in from a distance? 
\item How did the Solar System acquire ${}^{26}{\rm Al}$ and other radionuclides,
and is the amount it acquired common? 
\item {\bf Did the Sun form in a high-mass or low-mass star-forming region?}
\end{itemize}

To answer these questions, below we review astronomical observations, astrophysical
modeling, and planetary and meteoritic data.
We conclude that the Sun formed in a high-mass star-forming region with $N \sim 10^3 - 10^4$ stars.
Its disk of solids late ($> 5$ Myr) in its evolution is inferred to have been
$\approx 50$ AU, and its gas disk $\approx 50-90$ AU.
More likely than not, it was cut off from accretion from the molecular cloud
within a few $\times 10^5$ yr of its birth, but it cannot be ruled out that it
accreted gas for a few Myr, although it is not clear that gas would be compositionally
different.
The amount of ${}^{26}{\rm Al}$ it received appears typical for stars formed in
a high-mass star-forming region, acquired from a molecular cloud enriched over
the previous tens of Myr by massive stars.
The outskirts of the Orion Nebula, more than 2 pc from the center, appear to be a good analog environment.

\section{Evidence the Sun formed near massive stars}

\subsection{Astronomical evidence}

Planetary and astronomical observations suggest that the Sun
itself formed in a region with high stellar density and high UV flux,
both signposts of formation in a high-mass star-forming region.

\subsubsection{\it Structure of the Kuiper Belt}

The strongest evidence comes from the architecture of the Kuiper Belt,
especially the population of `detached' Kuiper Belt Objects (KBOs).
One of the major populations of KBOs is scattered objects, with large
semi-major axes $a$ and perihelia $q \approx 30$ AU.
These were evidently scattered by Neptune as it migrated outward.
{The timing of Neptune's migration is constrained to be $\approx 60$ Myr, by the sequence of events that led to the formation of the dwarf planet Haumea and its dynamical family \citep{NovielloEtal2022}.
It is also very likely that Neptune's migration was due} to the outer Solar System dynamical (`Nice') instability
\citep{TsiganisEtal2005},
which simulations of Solar System analogs suggest took place most likely
at 37-62 Myr \citep{deSousaEtal2020}.
{It is also likely that the Nice instability led to the Giant Impact that formed the Earth-Moon system, at about 60 Myr \citep{BarboniEtal2017}.}
A subset of about 200 of these {objects scattered by Neptune} have $a > 50$ AU and
detached perihelia $35 \, {\rm AU} < q < 50 \, {\rm AU}$ that require
them to have been gravitationally perturbed after they were scattered
\citep{KenyonBromley2004,MorbidelliLevison2004,BrasserEtal2006}.
Sedna, with $a = 506$ AU and $q = 76$ AU, is a similar but more extreme example.
Their radial distribution has been successfully modeled only by a stellar
encounter; the one simulation that matched observations involved a $0.17 \, M_{\odot}$
star passing 175 AU from the Sun, after Neptune's migration.
This would be improbable unless the Sun formed in a dense cluster.

For the stellar encounter creating the detached KBOs to be probable requires
the product of stellar density, $\eta$, and time exposed to this density, $T$,
to be $\eta \, T > 1 \times 10^4 \, {\rm pc}^{-3} \, {\rm Myr}$.
Assuming the encounter took place in the first 100 Myr of Solar System evolution, but after Neptune's migration at $\approx 50$ Myr, a lower limit $\eta > 200 \, {\rm pc}^{-3}$ is inferred.
(An interval of tens of Myr, comparable to the timescale of Neptune's migration, also is necessary to create sufficient numbers of detached scattered KBOs;  \citealt{NesvornyEtal2023}).
An upper limit $\eta \, T < 3 \times 10^{4} \, {\rm pc}^{-3} \, {\rm Myr}$
can be placed from the relatively low inclinations of dynamically cold KBOs
\citep{BatyginEtal2020}.
Assuming here $T \approx 100$ Myr, $\eta < 300 \, {\rm pc}^{-3}$ is inferred.
These limits are loose, based on stochastic probabilities, but suggest a stellar density $\approx 250 \, {\rm pc}^{-3}$, or $\approx 100 \, M_{\odot} \, {\rm pc}^{-3}$ (assuming an average stellar mass $0.4 \, M_{\odot}$; \citealt{Kroupa2001}). 
Just as importantly, this high stellar density must have lasted for up to 100 Myr (tens of Myr after Neptune's migration) to explain the Kuiper Belt.

Therefore, it is essential that the Sun formed in a cluster with a minimum number of sibling stars and a sufficient density to remain gravitationally bound for many tens of Myr, which is quite diagnostic. 
For example, \citet{ForbesEtal2021} have suggested the $\rho$ Ophiuchi is a good analog to the Solar System. 
We also find that the UV flux and stellar density there
are good matches to what we infer for the early Solar System, and stars in that region are likely to acquire live ${}^{26}{\rm Al}$; but the cluster is not likely to stay bound.
The $\rho$ Ophiuchi cluster has about 500 stars and a stellar density $\approx 90 \, M_{\odot} \, {\rm pc}^{-3}$ \citep{RatzenbockEtal2023, MiretRoigEtal2022, EvansEtal2009}; but integrating the trajectories of the stars in a Galactic potential, these will disperse to densities $< 12 \, M_{\odot} \, {\rm pc}^{-3}$ in 10 Myr, and $< 1 \, M_{\odot} \, {\rm pc}^{-3}$ in 100 Myr (Miret-Roig et al., in prep.).
The vast majority of stars similarly disperse from their birth clusters in $< 10$ Myr, and only a small fraction $\approx 10\%$ remain gravitationally bound to their siblings for longer times \citep{LadaLada2003, Adams2010}.
What determines whether a bound, open cluster forms is not clear, but
observational evidence suggests the mechanism works most efficiently in clusters with roughly $N > 2000$ stars \citep{LadaLada2003,HaoEtal2023} to
$N > 10^4$ stars \citep{Adams2010}.
The Orion Nebula, with 2200 stars within 2 pc \citep{MegeathEtal2012}, is generally considered likely to form a bound cluster, but just marginally so \citep{HillenbrandHartmann1998,KroupaEtal2001}.

\subsubsection{\it Truncation of the Disk}

A second line of evidence for the Sun's formation in a high-mass star-forming region
comes from the inferred size of the protoplanetary disk.
The orbits of cold classical KBOs, with low orbital inclinations and eccentricities
that reflect formation in the disk, extend only out to semi-major axes $\approx 47$ AU
\citep{JewittEtal1998,TrujilloBrown2001}, strongly suggesting KBOs did not form
beyond this distance.
Because the growth of KBOs is attributed to streaming instability \citep{McKinnonEtal2020}
and streaming instability effectively only concentrates particles with Stokes numbers
$> 10^{-3}$ \citep{LiYoudin2021}, this indicates a probable lack of ${\rm St} > 10^{-3}$
particles (i.e., particles larger than tens of $\mu{\rm m}$ in size) beyond about 47 AU.

{Two ideas exist to explain why protoplanetary disks might be truncated.}
One idea is tidal truncation of both gas and solids in the disk, due to the gravitational effects of a star passing at about 150 AU \citep{IdaEtal2000}.
The second is that external photoevaporation by UV radiation with
$G_0 \sim 10^3$ also could truncate the disk at 50-100 AU \citep{HollenbachAdams2004}.
Both mechanisms would demand formation in a high-mass star-forming region.
{For a given stellar density, studies show that disks are much more likely to be truncated by UV photoevaporation from massive stars than by dynamical interaction with passing stars \citep{WinterEtal2018}.}

Although KBOs trace the primordial distribution of solid matter only, the gas density also can be inferred, 
{with the recognition that gas and particles are decoupled.
In recent years, it has been understood} that the gas disk
extended beyond 50 AU, but that large particles had drifted radially inward due to aerodynamic drag by the time KBOs formed.
For example, using standard formulas
\citep{DeschEtal2017}, 
and assuming a sound speed $c_{\rm s} \approx 0.3 \, {\rm km} \, {\rm s}^{-1}$ and
typical disk properties, only particles with Stokes numbers ${\rm St} < 10^{-3}$
would have drift speeds $v_{r,{\rm d}} < 5 \, {\rm cm} \, {\rm s}^{-1}$ and avoid
spiralling inward tens of AU within 5 Myr, the typical timescale for formation of KBOs
\citep{BiersonNimmo2019}.
This implies much reduced masses of submillimeter particles beyond a radius very close
to the observed 47 AU.
This would make the solar nebula consistent with other evolved protoplanetary disks,
which are commonly observed to have gas emission that extends up to a factor of 2 larger
radii than millimeter emission from solids \citep{AnsdellEtal2018}.
The Sun's gas disk probably extended out to at least 50 AU, and possibly closer to 90 AU.

The size of the Sun's protoplanetary disk argues for formation in an intermediate-mass star-forming region.
\citet{MarchingtonParker2022} compiled literature data about disk sizes and masses, and $G_0$,
across various star-forming regions, finding disks are generally smaller in more massive
star-forming regions, largely due to external photoevaporation.
In low-mass star-forming regions, $G_0$ is low and disks are large:
in the several Myr-old Taurus-Auriga region, $G_0 \sim 1$ and the median radius of dust emission,
 $r_{\rm d}$, is about 50 AU; and
in the 6 Myr-old \citep{Luhman2020b} Lupus region, $G_0 \sim 4$ 
\citep{GalliEtal2020,CleevesEtal2016}, and the median $r_{\rm d} \approx 100$ AU
\citep{TazzariEtal2017,AnsdellEtal2018}.
In the 3--10 Myr-old \citep{RatzenbockEtal2023, MiretRoigEtal2022} Upper Sco region, and in the adjacent $\rho$ Oph cloud, $G_0 \approx 50$
\citep{TrapmanEtal2020,Luhman2020a,ParkerEtal2012} and the median $r_{\rm d} \approx 60$ AU.
Within the central parsec or so of the 2.5 Myr-old ONC \citep{JeffriesEtal2011,DaRioEtal2010}, $G_0 \sim 10^4$
\citep{HillenbrandHartmann1998} and the median $r_{\rm d} \approx 15$ AU.
Finally, the majority of disks in the Trumpler 14 cluster in the Carina Nebula, which see
$G_0 \sim 10^6$, have $r_{\rm d} < 3$ AU \citep{MesaDelgadoEtal2016}.
Formation in the Carina Nebula is probably excluded, as is formation in the central parts of the Orion Nebula, where $< 5\%$ of dust disks are as large as 47 AU \citep{MarchingtonParker2022}.
The Sun's disk size is more compatible with environments with lower $G_0$, like Lupus or Taurus-Auriga, or Upper Sco, where disks see a range of $G_0$ from 5 to 500 \citep{TrapmanEtal2020}, or parts of the Orion Nebula $> 2$ pc from the center.
The upper limit on $G_0$ is $> 10^3$ but $< 10^4$.

One last aspect of externally photoevaporated disks is more compatible with intermediate
$G_0$ rather than a Taurus-like $G_0 \approx 1$.
The {\it Stardust} samples of the Wild 2 comet include refractory inclusions like {\it Inti}
thought to have formed inside 1 AU, then transported outward in the disk to the comet-forming
region \citep{ZolenskyEtal2006}.
Because the gas flow in disks is inward inside the transition radius, $r_{\rm T}$, and outward
for $r > r_{\rm T}$, this requires $r_{\rm T}$ to remain at a few AU even late in disk evolution,
in contrast to self-similar models of isolated, viscously evolving disks, which predict outward
movement of $r_{\rm T}$ to $> 10$ AU after a few Myr \citep{HartmannEtal1998}.
Even modest amounts of external photoevaporation ($G_0 \sim 30$) can keep $r_{\rm T}$ fixed at about 3 AU, but not $G_0 \sim 1$ \citep{Desch2007,KalyaanEtal2015,DeschEtal2018}.

\vspace{0.1in}

\subsubsection{\it Jupiter's Noble Gas Abundances}

A third line of evidence for high UV fluxes comes from Jupiter's noble gas abundances.
The abundances of various species X relative to H and solar abundances,
$({\rm X}/{\rm H}) / ({\rm X}/{\rm H})_{\odot}$, have been measured
in Jupiter's atmosphere:
He, Ne, Ar, Kr, Xe, C, N, O, and S by the {\it Galileo} entry probe Neutral Mass
Spectrometer; P by the {\it Galileo} orbiter Infrared Mapping Spectrometer;
and O by the {\it Juno} microwave radiometer \citep{LiEtal2020}.
Compared to solar abundances \citep{AsplundEtal2009},
He and Ne are slightly depleted, certainly due to selective sequestration into
liquid helium droplets that sink to Jupiter's core 
\citep{WilsonMilitzer2010}, 
but the other species are all enhanced by the same factor, $\approx 3$
\citep{GuillotHueso2006,MongaDesch2015,LiEtal2020}.

Enhancements of the noble gases, especially Ar, are difficult to explain,
as they fractionate from ${\rm H}_{2}$/He gas only {when Ar enters ice, which requires deposition of water vapor as amorphous ice, at temperatures $< 25$ K
\citep{BarNunEtal1988}.}
The only successful model for these uniform enhancements is that H (and He and Ne),
but not Ar and other elements, were lost from the solar nebula before Jupiter
accreted most of its gas (i.e., by about 5 Myr; \cite{DeschEtal2018}).
If two-thirds of the just the H$_2$, He and Ne were lost, then all other species
would appear uniformly enhanced relative to H, by the same factor of 3.
\citet{MongaDesch2015}, building on the model of \citet{GuillotHueso2006},
showed that if photoevaporation removes H$_2$ gas from the outer edge of the disk at a location
where the disk temperature is $< 25$ K (i.e., beyond 50 AU), then Ar could potentially
be trapped in large amorphous ice grains that could spiral back in the planet-forming
region.

The \citet{MongaDesch2015} model is consistent with $G_0 \approx 300$.
A lower limit of roughly $G_0 > 30$ can be placed by the need in the model for UV to photodesorb enough water vapor to recondense and trap Ar and noble gases. 
An upper limit of about $G_0 < 3000$ can be placed by the need for gas loss at the outer edge of the disk to take place where temperatures are cold enough (roughly 25 K) to trap Ar, which requires the outer disk edge to remain at 50-100 AU.
\vspace{0.1in}

\subsection{Meteoritic evidence}

Further tests of whether or not the Sun formed in a high-mass star-forming region
could come from astrophysical modeling combined with meteoritic data, especially
data pertaining to the mass-independent fractionation of oxygen and sulfur isotopes,
and short-lived radionuclides.

\subsubsection{\it Oxygen Isotopes}

\begin{figure}[ht]
\centering
\includegraphics[width=0.95\textwidth]{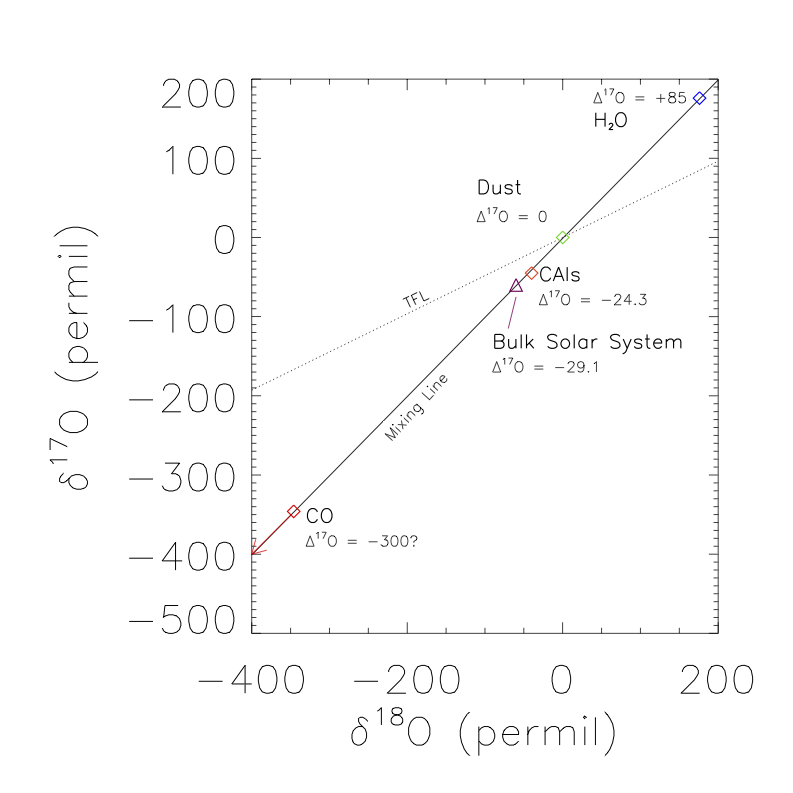}
\caption{
Three-isotope plot on which terrestrial samples are typically found at $\delta^{17}{\rm O} = \delta^{18}{\rm O} = 0\permil$ or along the Terrestrial Fractionation Line (TFL) with slope 0.52. Planetary materials are typically found near the terrestrial value, but the Sun is found on a mixing line with slope 1.0, at $\delta^{17}{\rm O} = \delta^{18}{\rm O} \approx -60\permil$ and $\Delta^{17}{\rm O} = -29.1\permil$; CAIs are found on the same mixing line, at $\delta^{17}{\rm O} = \delta^{18}{\rm O} \approx -50\permil$ and $\Delta^{17}{\rm O} = -24.3\permil$.  The mixed components are hypothesized also to fall on this line: isotopically heavy water ($\Delta^{17}{\rm O} \approx +85\permil$), normal dust ($\Delta^{17} \approx 0\permil$) and light CO ($\Delta^{17}{\rm O} < -300\permil$).
\label{fig:oxygen}
}
\end{figure}

{A very useful tool in planetary science is the ``three-isotope" plot of oxygen isotopic compositions, like the one displayed in Figure~\ref{fig:oxygen}.
The vertical axis is $\delta^{17}{\rm O} = ({}^{17}{\rm O}/{}^{16}{\rm O}) / ({}^{17}{\rm O}/{}^{16}{\rm O})_{\rm SMOW} -1$ and the horizontal axis is $\delta^{18}{\rm O} = ({}^{18}{\rm O}/{}^{16}{\rm O}) / ({}^{18}{\rm O}/{}^{16}{\rm O})_{\rm SMOW} -1$, where the terrestrial standard, Standard Mean Ocean Water, has $\delta^{17}{\rm O} =$ $\delta^{18}{\rm O} = 0\permil$ (where $\permil$ is parts per thousand).}
Almost all chemical processes fractionate oxygen isotopes in a mass-dependent way, along a line with slope 0.52. 
The vertical displacement of another sample from this line is $\Delta^{17}{\rm O} = \delta^{17}{\rm O} - 0.52 \, \delta^{18}{\rm O}$.
Essentially all samples on Earth fall along a single line of slope 0.52 and
$\Delta^{17}{\rm O} = 0\permil$.
Samples from another asteroid will also array along a parallel line of slope 0.52,
with a common, distinct $\Delta^{17}{\rm O}$; for example, $\Delta^{17}{\rm O} = -0.3\permil$
for all meteorites from the asteroid 4 Vesta \citep{IrelandEtal2020}.

Surprisingly, calcium-rich, aluminum-rich inclusions (CAIs) contain minerals arraying along a line of slope 1.0 in a three-isotope plot, {some extending to} planetary-like values $\Delta^{17}{\rm O} \approx 0\permil$, {while most CAIs plot at} $\Delta^{17}{\rm O} < -24\permil$ \citep{IrelandEtal2020}.
This is interpreted as a mixing line between planetary (silicate) materials and some other,
${}^{16}{\rm O}$-rich (or ${}^{17,18}{\rm O}$-poor) reservoir.
The Sun lies at $\Delta^{17}{\rm O} \approx -29\permil$, close to CAIs in the three-isotope plot,
as inferred from solar wind measurements by the {\it Genesis} mission \citep{McKeeganEtal2011}.
The process that created that other reservoir is distinct from normal chemical processes, and is
called ``mass-independent fractionation" (MIF).

Practically the only widely accepted model for the MIF is isotopically selective photodissociation of CO gas in the protoplanetary disk or molecular cloud, by UV radiation from either the Sun or a nearby massive star
\citep{YurimotoKuramoto2004,LyonsYoung2005,Young2007}.
Because CO is photodissociated by first exciting the molecule to a distinct (but unbound) energy level, ${\rm C}^{16}{\rm O}$, ${\rm C}^{17}{\rm O}$, and ${\rm C}^{18}{\rm O}$ are photodissociated by photons of different UV wavelengths.
Because ${}^{16}{\rm O}$ is so much more abundant, the ${\rm C}^{16}{\rm O}$-dissociating photons are absorbed close to the edge of the disk or molecular cloud, an effect called ``self-shielding".
Gas near the edge also is likely to escape in a photoevaporative flow.
For practical purposes, the material incorporated into meteorites comes from deep within the disk or cloud, in which ${\rm C}^{17}{\rm O}$ and ${\rm C}^{18}{\rm O}$ are
equally photodissociated.
This leaves the remaining CO richer in ${}^{16}{\rm O}$, with
$\delta^{17}{\rm O} \approx \delta^{18}{\rm O}$ and very negative $\Delta^{17}{\rm O}$.
The liberated O reacts with hydrogen to form ${\rm H}_{2}{\rm O}$ uniformly enriched in
${}^{17}{\rm O}$ and ${}^{18}{\rm O}$, with
$\delta^{17}{\rm O} \approx \delta^{18}{\rm O}$ and very positive $\Delta^{17}{\rm O}$.
Any mixture sampling oxygen from the starting composition of silicates, CO gas and ${\rm H}_{2}{\rm O}$
will array along a line of slope 1.0 in a three-isotope plot.
This model is strongly supported by the ``cosmic symplectite" material in the Acfer 094 chondrite,
containing minerals like magnetite that are products of aqueous alteration, that clearly
exchanged O isotopes with ${\rm H}_{2}{\rm O}$;
it has $\delta^{17}{\rm O} \approx \delta^{18}{\rm O}$
$\approx +180\permil$ and $\Delta^{17}{\rm O} \approx +85\permil$
\citep{SakamotoEtal2007}.
The solar nebula appears to have had isotopically light CO, isotopically heavy water, and
silicates with intermediate isotopic composition near $\delta^{17}{\rm O} \approx \delta^{18}{\rm O} \approx 0\permil$.

{\Steve Strong support for self-shielded photodissociation of CO comes from astronomical observations of CO absorption in some protostellar systems.
\citet{SmithEtal2009} measured the values of $\delta^{17}{\rm O}$ and $\delta^{18}{\rm O}$ in CO gas in the protoplanetary disk in the VV CrA system and found an excess of ${}^{16}{\rm O}$ on the order of $\Delta^{17}{\rm O} \approx -200\permil$, as expected if CO gas was photodissociated in an isotopically selective way.
Unfortunately, such observations leave open the question of whether the CO was photodissociated by UV radiation from the central protostar or from nearby OB stars, and whether the CO was photodissociated in the disk or during an earlier molecular cloud stage.
}

{\Steve 
If irradiation in the disk could be ruled out, then the irradiation would have to have taken place in the molecular cloud, due to nearby OB stars.
Models of self-shielding in the disk have been constructed \cite[e.g.,][]{Young2007,LyonsEtal2009}, although the need to create distinct reservoirs, well-mixed across the entire disk, before the formation
of many mineral phases in CAIs, is problematic (see \citealt{AlexanderEtal2017,KrotEtal2020}).
Such models rely on large values of the the turbulence parameter, $\alpha \sim 10^{-2}$, so that the isotopic signature produced in the outer disk can be transmitted to the inner disk in $< 3 \times 10^4$ years \citep{Young2007,LyonsEtal2009}.
Whether such a high $\alpha$ is reasonable or not is debated.
It is difficult to argue from first principles what the value should be because the physical mechanism of transport in disks has not been identified definitively.
While many observations argue for $\alpha \ltsimeq 10^{-4}$ in the outer portions of many protoplanetary disks \citep{YoudinLithwick2007,PinteEtal2016,FlahertyEtal2020}, there is observational support for high values in some disks \citep{SenguptaEtal2024}.
Models of the outer regions of the Sun's disk in particular argue for a low value $\alpha \sim 10^{-5}$ or else there would be rapid mixing between the ``NC" inner-disk reservoir with the ``CC" outer disk reservoir \citep{DeschEtal2018}.
As well, the values of $\alpha$ leading to mixing of an outer disk $\Delta^{17}{\rm O}$ signature with the inner disk would also lead to rapid transport of angular momentum and high mass accretion rates.
The coefficient of turbulent mixing, ${\cal D} \sim r^2 / t_{\rm mix}$ is comparable (${\cal D} \sim \nu$) to the turbulent viscosity $\nu = \alpha C^2 / \Omega$.
Because the mass accretion rate is $\dot{M} \approx 3\pi \Sigma \nu$, the mass-loss timescale $(\pi r^2 \, \Sigma) / \dot{M}$ therefore is necessarily comparable to the mixing timescale $t_{\rm mix}$.
Rapid mixing of the $\Delta^{17}{\rm O}$ signature from the outer disk to the inner disk before CAI formation would also imply loss of the disk on comparable timescales $\sim 10^5$ years, not several Myr. 
Then again, models for the NC/CC dichotomy exist that invoke a rapid evolution of the solar nebula and high values of $\alpha$ that might allow rapid mixing \citep{LiuEtal2022}.
There are significant challenges to models in which self-shielding exists in the disk, but such models have not been ruled out.
}

{\Steve In contrast,} isotopically selective photodissociation of CO in the molecular cloud is a more established idea \citep{BallyLanger1982} and the basis for models of MIF of oxygen isotopes in the
Solar System in particular \citep{YurimotoKuramoto2004}.
Unfortunately, such models do not make firm predictions of $G_0$, but values $\gg 1$ presumably are needed to create $\delta^{16}{\rm O}$-poor H$_2$O deep in the molecular cloud.

\subsubsection{\it Sulfur Isotopes}

Recently \citet{VacherEtal2021} analyzed magnetite-pentlandite minerals in the same cosmic symplectite material in Acfer 094, for S isotopes.
Like oxygen, S isotopes should fractionate according to a slope-1/2 line in \linebreak
$\delta^{33}{\rm S} = ({}^{33}{\rm S}/{}^{32}{\rm S}) / ({}^{33}{\rm S}/{}^{32}{\rm S})_{\rm CDT} - 1$ vs.\
$\delta^{34}{\rm S}$, or a slope-2 line in $\delta^{36}{\rm S}$ vs.\ $\delta^{34}{\rm S}$
space (here CDT is the terrestrial standard, Canyon Diablo Troilite).
The deviations of a sample from these lines can be labeled $\Delta^{33}{\rm S}$ or $\Delta^{36}{\rm S}$.
Almost all samples on Earth---and indeed, throughout most of the Solar System---have
$\Delta^{33}{\rm S} \approx 0\permil$ and $\Delta^{36}{\rm S} \approx 0\permil$; but
the cosmic symplectite is characterized by
$\Delta^{33}{\rm S} \approx +4\permil$ and $\Delta^{36}{\rm S} \approx -6\permil$,
yielding a ratio $\Delta{}^{36}{\rm S}/\Delta^{33}{\rm S} = -1.58 \pm 0.66$.

\citet{VacherEtal2021} interpreted this as arising from photodissociation of
${\rm H}_{2}{\rm S}$ in the molecular cloud.
The details of photodissociation of ${\rm H}_{2}{\rm S}$ differ from those of CO, but lead
to the same result: the different isotopes of S are liberated from gas-phase ${\rm H}_{2}{\rm S}$,
as elemental ${\rm S}^{0}$, at different rates.
As long as the temperature is $< 370$ K, the ${\rm S}^{0}$ recondenses on dust grains before
converting back to ${\rm H}_{2}{\rm S}$, imparting a MIF isotopic signature in S in solids.
The resultant ratio $\Delta{}^{36}{\rm S}/\Delta^{33}{\rm S}$ in the solids is sensitive
to the spectrum of the UV radiation.
UV radiation from T Tauri stars like the early Sun is dominated by Ly$\alpha$ photons, and
photodissociation by this spectrum results in
 ${\rm S}^{0}$ with $\Delta{}^{36}{\rm S}/\Delta^{33}{\rm S} \approx -2.73 \pm 0.73$
 \citep{ChakrabortyEtal2013}, incompatible with the isotopic signature.
Photodissociation by UV radiation from OB stars, or the interstellar radiation field,
result in ${\rm S}^{0}$ with
$\Delta{}^{36}{\rm S}/\Delta^{33}{\rm S} \approx -1.53 \pm 0.67$ or $-1.75 \pm 0.71$,
repectively, matching the cosmic symplectite.

{\Steve 
In similarity to self-shielding of CO, the isotopic anomalies themselves ($\Delta^{17}{\rm O}$, $\Delta^{34}{\rm S}$, $\Delta^{36}{\rm S}$) do not directly constrain the strength $G_0$ of the UV field. 
Self-shielding of H$_2$S in the protoplanetary disk rather than molecular cloud also probably cannot be ruled out at this time.
However, unlike the case of CO, photodissociation of H$_2$S seems to demand a UV flux from nearby OB stars, which therefore constrains the Sun's birth environment.}

{\Steve
The photodissociation of CO and H$_2$S together also do place some indirect limits on the UV flux and $G_0$, through the gas temperatures that are implied.
}
CO must be in the gas phase where irradiation takes place, implying temperatures $> 50$ K; but ${\rm H}_{2}{\rm O}$ must condense, suggesting temperatures $< 170$ K, roughly.
Likewise, ${\rm H}_{2}{\rm S}$ must be in the gas phase, implying temperatures $> 70$ K; but $S^0$ must condense before back-reacting to ${\rm H}_{2}{\rm S}$, implying temperatures $< 370$ K \citep{VacherEtal2021}.
These temperatures, {\Steve $\approx 70-170$ K, are quite characteristic of photodissociation regions (PDRs) with intermediate UV flux.}
In the PDR models of  \citet{AdamsEtal2004}, these temperatures arise at various optical depths into gas with $G_0$ in the range of 300 to $3 \times 10^4$. 
{\Steve They did not present results for lower $G_0$, but a}
lower limit to $G_0 > 10$ or so also must exist: most molecular clouds and filaments exposed to smaller $G_0$ are overwhelmingly associated with very cold temperatures $< 20$ K (e.g., \citet{HacarEtal2023}), at which CO and ${\rm H}_{2}{\rm S}$ would be in the solid phase (although desorption of molecules from grains by GCR bombardment complicates this picture).

\subsubsection{\it Short-lived radionuclides}

For a long time, the existence of short-lived ($t_{1/2} < $ tens of Myr) radionuclides
(SLRs) like ${}^{26}{\rm Al}$ in the solar nebula was taken as demanding formation
very near a supernova \citep{CameronTruran1977,Adams2010}.
The supernova was presumed to inject ${}^{26}{\rm Al}$ and other SLRs either into
the Sun's protoplanetary disk or into its molecular cloud, just prior to collapse of the
Sun's cloud core.
Models requiring such proximity of a Wolf-Rayet star or supernova to the Sun
tend to require higher stellar densities $\gg 10 \, M_{\odot} \, {\rm pc}^{-3}$
\citep{PfalznerVincke2020,ParkerEtal2023}
and larger number of stars (e.g., $N > 2500$; \citealt{PortegiesZwart2019}) in the Sun's
birth cluster, but even then the relative timing of events would be improbable
\citep{OuelletteEtal2010}.
It seems more likely instead that the Sun and its protoplanetary disk inherited their SLRs
from the molecular cloud, which was itself enriched by Wolf-Rayet winds and supernovae,
by virtue of being in a high-mass star-forming region.

Strong evidence for the inheritance of SLRs from the molecular cloud is provided
by their homogeneity in meteoritic samples, especially bulk abundances of achondrites.
\citet{DeschEtal2023a} and \citet{DeschEtal2023b} compiled and analyzed the times of 
formation of 14 bulk achondrites
according to radiometric dating using the Pb-Pb, Al-Mg, Mn-Cr, and Hf-W chronometers.
Across these systems, 38 times of formation are made concordant in a statistically
rigorous sense ($\chi_{\nu}^{2} = 1.1$), assuming values for only four parameters
(initial abundances of ${}^{53}{\rm Mn}$ and ${}^{182}{\rm Hf}$, a half-life of 3.8 Myr
for ${}^{53}{\rm Mn}$, and a Pb-Pb age of the Solar System of 4568.4 Myr).
This strongly supports homogeneity in the solar nebula of the SLRs ${}^{26}{\rm Al}$,
${}^{53}{\rm Mn}$, and ${}^{182}{\rm Hf}$, which therefore must have been inherited
from the molecular cloud.

The abundances of {\it all} the known SLRs in the early Solar System of nucleosynthetic origin
(${}^{41}{\rm Ca}$, ${}^{36}{\rm Cl}$,  ${}^{26}{\rm Al}$,  ${}^{60}{\rm Fe}$, ${}^{53}{\rm Mn}$,
${}^{107}{\rm Pd}$, ${}^{182}{\rm Hf}$,  ${}^{247}{\rm Cm}$, ${}^{129}{\rm I}$),
as well as longer-lived radionuclides (${}^{92}{\rm Nb}$, ${}^{244}{\rm Pu}$,
${}^{146}{\rm Sm}$, ${}^{235}{\rm U}$, ${}^{238}{\rm U}$) appear remarkably consistent with
the predictions of a simple box model in which the Sun formed in a spiral arm of the Galaxy, from molecular clouds continuously enriched by inputs from massive stars
(\cite{Jacobsen2005}, \cite{HussEtal2009}, \cite{Young2014}, \cite{Young2020}; see review by 
\cite{DeschEtal2023c}).
In essence, massive stars forming in the spiral arms rapidly evolve and eject material
as Wolf-Rayet winds and supernovae before leaving the spiral arm, contaminating molecular clouds,
which are the sites of subsequent star formation.
Because molecular clouds are gravitationally drawn into the spiral arms, the SLR enrichments
do not readily leave the spiral arms, allowing some steady-state buildup over many $\times 10^8$ yr
within these structures \citep{FujimotoEtal2018,FujimotoEtal2020}.
While very short-lived SLRs (${}^{41}{\rm Ca}$, ${}^{26}{\rm Al}$) must be much less abundant
in steady state than longer-lived ones, their abundances, like the abundances of all 14
radionuclides above, are exactly consistent with the predictions of the model, assuming
the material entering molecular clouds is overwhelmingly from Wolf-Rayet winds and not from
supernovae.
This is surprising but justifiable: massive stars evolve to eject Wolf-Rayet winds within $< 4$ Myr, while they are still near their birth clouds, whereas most stars explode as supernovae only after $> 10-20$ Myr, by which time they have dispersed far from their birth clouds.

{At one point, a high abundance ${}^{60}{\rm Fe}/{}^{56}{\rm Fe} \sim 10^{-6}$ was inferred to have existed in the solar nebula, which would have practically demanded injection of material from a nearby supernova \citep{TachibanaEtal2006}.
Subsequent analyses have shown how these values were surely overestimated \citep{OglioreEtal2011,TelusEtal2016}.
It is now understood that ${}^{60}{\rm Fe}$ did exist in the solar nebula, but at levels ${}^{60}{\rm Fe}/{}^{56}{\rm Fe} \sim 10^{-8}$ \citep{TangDauphas2012}. 
These levels are consistent with the levels inferred from gamma ray observations to exist in the Galaxy and in molecular clouds \citep{FujimotoEtal2018,DeschEtal2023c}.
While the Sun probably formed in a region with supernovae, the evidence from ${}^{60}{\rm Fe}$ does not demand it formed near ($< 1$ pc) a contemporaneous supernova.
}

Molecular clouds in spiral arms also should contain ${}^{10}{\rm Be}$ at levels consistent
with the values in the solar nebula.
The abundance of ${}^{10}{\rm Be}$ in the solar nebula was a remarkably uniform
${}^{10}{\rm Be} / {}^{9}{\rm Be} \approx 7.1 \times 10^{-4}$, indicating inheritance
of ${}^{10}{\rm Be}$ from the Sun's natal molecular cloud \citep{DunhamEtal2022}, 
as with the other SLRs.
Live ${}^{10}{\rm Be}$ is produced by spallation of oxygen and other nuclei, by
Galactic cosmic rays (GCRs).
The solar nebula abundance was thought to be too high to have been inherited from the Sun's
molecular cloud \citep{TatischeffEtal2014}, assuming a GCR flux based on observations of
the Sun's current neighborhood, which is relatively quiescent.
But molecular clouds in spiral arms see higher GCR flux, as GCRs are accelerated in
supernova shocks, and supernovae are concentrated in spiral arms.
Allowing for the higher GCR fluxes in spiral arms 4.57 Gyr ago, a ratio
${}^{10}{\rm Be}/{}^{9}{\rm Be} = 7.1 \times 10^{-4}$ is a very probable outcome
\citep{DunhamEtal2022,DeschEtal2023c}.

While SLRs like ${}^{26}{\rm Al}$ were once taken as the smoking gun for birth of
the Sun near a massive star, the story is probably more subtle.
The abundances of the SLRs are all explained by inheritance from a molecular cloud in
a spiral arm, enriched in material from supernovae and especially Wolf-Rayet winds.
This places the Sun more definitively than ever in the general vicinity of a high-mass star-forming region, but not necessarily less than a parsec from a supernova. 
Acquisition of live ${}^{26}{\rm Al}$ from previous generations of supernovae and massive stars would seem readily achieved in regions like Upper Scorpius \citep{KrauseEtal2018,RatzenbockEtal2023}
or the Orion Nebula 
\citep{FoleyEtal2023}.

\section{The Sun's formation}

{Based on the arguments presented here, summarized in Table 1, we place the Sun's formation in a molecular cloud in a spiral arm, contaminated by previous generations of star formation.
The Sun formed in a bound stellar cluster that maintained densities of several hundred stars per cubic parsec, for about 100 Myr.
It experienced a UV flux much higher than is found in the interstellar medium outside star-forming regions.}
\begin{table}[h!]
\centering
\begin{tabular}{| p{1.1in} | p{1.05in} | p{2.48in} |}
\hline
{\bf Parameter} & {\bf Constraint} & {\bf Reason} \\
\hline
\hline
Cluster size & $N \, \gtsimeq \, 2000$ stars & Must form bound cluster. \\
\hline
Cluster lifetime & $T \approx 100$ Myr & Scattered KBOs' perihelia must be raised after scattering by Neptune migrating at $\approx 60$ Myr. \\
\hline
Stellar density & $\eta \approx 200$ - $300 \, {\rm pc}^{-3}$ & A stellar encounter must raise perihelia of scattered KBOs, but dynamically cold KBOs must not be over-excited. \\
\hline
UV Flux & $30 \, \ltsimeq \, G_0 \, \ltsimeq \, 3000$ & Jupiter's noble gas abundances; disk truncated at $\approx 100$ AU; outward gas flow beyond a few AU. \\
\hline 
Galactic star-forming environment & Molecular cloud in a spiral arm, enriched by Wolf-Rayet stars &
Abundances of short-lived radionuclides like 
${}^{26}{\rm Al}$, ${}^{10}{\rm Be}$, ${}^{53}{\rm Mn}$, ${}^{60}{\rm Fe}$, etc. \\
\hline
\end{tabular}
\caption{Constraints on the Sun's star-forming environment.}
\end{table}

\begin{figure}[ht]
\centering
\includegraphics[width=0.95\textwidth]{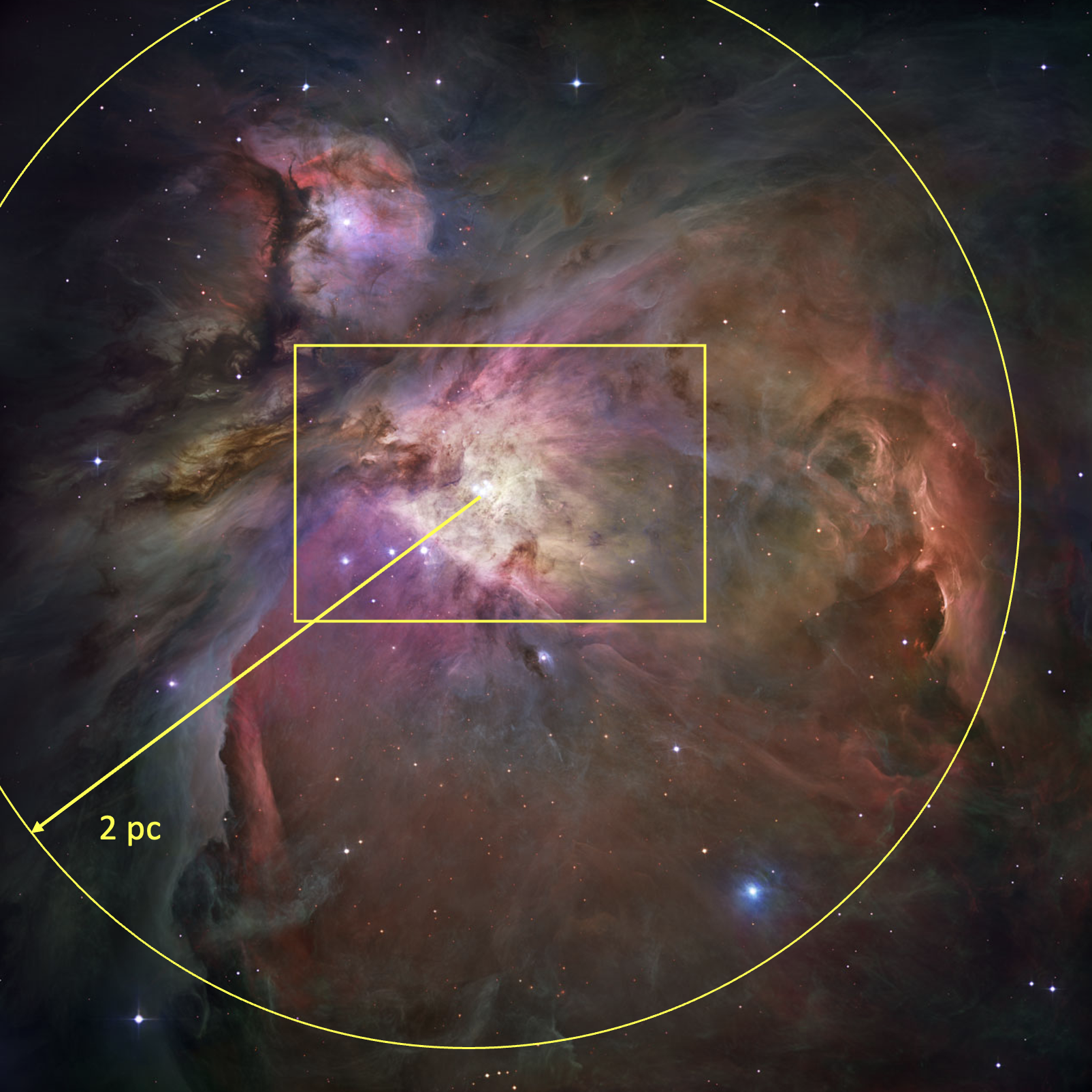}
\caption{{\it Hubble Space Telescope} optical ({\it Advanced Camera for Surveys}) image of the Orion Nebula. The field of view is 30.0 by 30.0 arcminutes, roughly 4.0 pc by 4.0 pc. The Trapezium cluster and $\theta^1$ Ori C are apparent at the center of the superimposed circle, which has a radius of 15 arcminutes, or 2 pc. The yellow rectangle shows the approximate field of view of the image in Figure~\ref{fig:orionjwst}.
(Credit: NASA, ESA, M. Robberto (Space Telescope Science Institute / ESA) and the Hubble Space Telescope Orion Treasury Project Team.)
\label{fig:orionhst}
}
\end{figure}

\begin{figure}[ht]
\centering\includegraphics[width=0.95\textwidth]{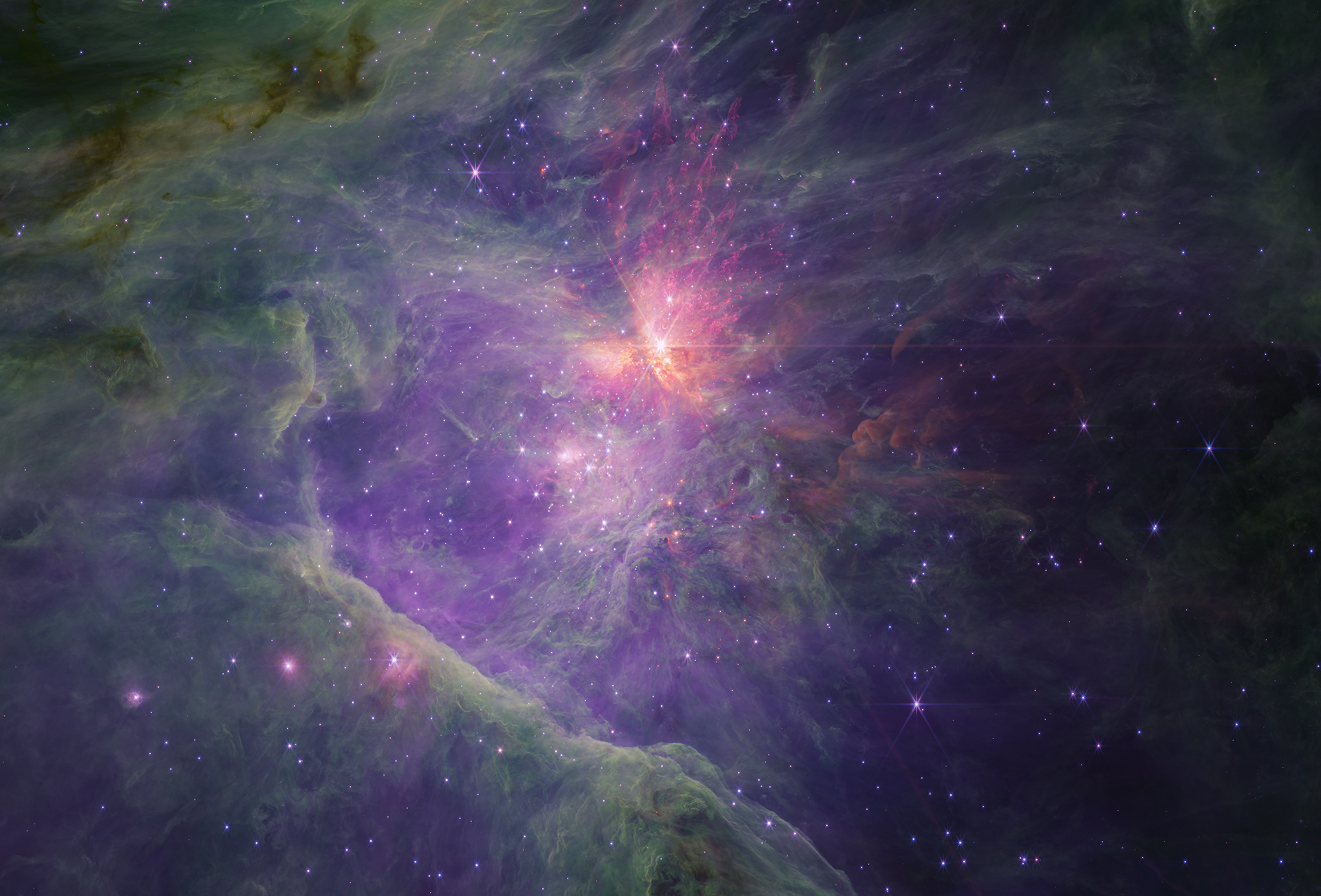}
\caption{{\it James Webb Space Telesope} {\it NIRCam} near-infrared image of the Orion Nebula. The field of view is 7.5 by 10.9 arcminutes, roughly 0.85 pc by 1.25 pc, and corresponds to the rectangle in Figure~\ref{fig:orionhst}.
Emission from ionized gas at the center is colored purple, while greens and browns denote emission from molecular gas and dust.  Below and to the left of the very bright BN-KL object (seen in reddish pink) is the Trapezium cluster including $\theta^1$ Ori C. 
(Credit: NASA/ESA/CSA/STScI, K. Pontoppidan, A. Pagan.)
\label{fig:orionjwst}
}
\end{figure}

The weight of this astronomical evidence places the Sun's formation in an environment like the outskirts of the Orion Nebula, perhaps 2 pc from the central Orion Nebula cluster and $\theta^1$ Ori C.
Figure~\ref{fig:orionhst} shows an optical image of the Orion Nebula taken by the {\it Hubble Space Telescope}, superimposed with a circle 15 arcminutes (2~pc at 450~pc) in radius, centered on $\theta^1$ Ori C. 
We consider the Sun to have formed in a region 
like those depicted in the left or upper corners of the figure, at the distance of the yellow circle. 
Although the density of stars there is not as great as in the center of the cluster, it is still substantial. \textcolor{black}{For reference, the Orion Nebula half-mass radius is $\lesssim 1$~pc \citep{MegeathEtal2016}}. In Figure~\ref{fig:orionjwst} we also show a recent near-infrared image taken by the NIRCam instrument on the {\it James Webb Space Telescope}  \citep{McCaughreanPearson2023}.
This image only depicts the central region ($< 2$ pc from the cluster center), but better conveys the structure of the nebulosity. 

Many aspects of the Orion Nebula match those inferred for the Solar System's early evolution.
As reviewed above, the Sun's molecular cloud and protoplanetary disk must have been exposed to a UV flux with $G_0 \approx 30 - 3000$, based on the need to create isotopically fractionated reservoirs of oxygen and sulfur in the molecular cloud, based on the truncation of the Sun's solid disk at $\approx 47$ AU (edge of the classical Kuiper Belt), based on the need for photoevaporation to shape the dynamics of the gas disk, and based on the need to remove hydrogen by photoevaporation to explain the abundances of Jupiter's atmosphere. 
The UV flux in the Orion Nebula scales with distance $r$ from the center of the cluster ($\theta^1$ Ori C) roughly as 
$G_0 \approx 5000 \, (r / 1 \, {\rm pc})^{-2}$ \citep{StorzerHollenbach1999}. Therefore, the Sun's birth place is consistent with environments at distances larger than about {1.3} pc from the Orion Nebula center, but not beyond {13} pc.
{Formation in the central 1 pc of a region like the Orion Nebula can be ruled out, as the UV flux and stellar density would be too high.}
The Sun's protoplanetary disk must have been exposed to a stellar density $\approx 100 \, M_{\odot} \, {\rm pc}^{-3}$ for $\approx 100 \, {\rm Myr}$.
The average stellar density in the Orion Nebula within 2 pc of the center is $\approx 100 \, M_{\odot} \, {\rm pc}^{-3}$, and it is likely to form a bound cluster, maintaining these stellar densities for 100 Myr \citep{HillenbrandHartmann1998}.
The Orion {Molecular Cloud} is expected to have already experienced 10-20 supernovae (some preceded by Wolf-Rayet winds) in the last 12 Myr \citep{Bally2008}, and the region is associated with gamma-ray emission from the decay of gas-phase ${}^{26}{\rm Al}$ 
{at a level similar to the solar nebula abundance of ${}^{26}{\rm Al}$}
\citep{Diehl2002}.
This highlights the likelihood of stars in regions like this incorporating live radionuclides like ${}^{26}{\rm Al}$.


Formation of the Sun in such an environment has important implications for the early evolution of the solar nebula.
If the Sun formed in a region like the Orion Nebula, its formation may have been triggered by the nearby massive stars, although the mechanism is probably not as simplistic as a cloud core triggered to collapse by a supernova shock.
A detailed history of triggered star formation has recently been worked out for the Scorpius-Centaurus region \citep{KrauseEtal2018,RatzenbockEtal2023}.
There, star formation is sequential, has been ongoing for the past 20 Myr, and has propagated inside-out towards multiple directions. 
 \citet{KrauseEtal2018} advocate that the Upper Sco molecular cloud was triggered to collapse by the surrounding superbubble of hot, X-ray-emitting gas created by previous generations of supernovae and massive star (Wolf-Rayet) winds.
\citet{RatzenbockEtal2023} refined this idea to suggest that the superbubble
triggered collapse and star formation in gas arranged along filaments, resulting in chains of star clusters.
Similarly, in the Orion Nebula, the center of ${}^{26}{\rm Al}$ gamma-ray emission appears coincident with a center of expansion of stars, implying one or more stellar explosions at that point about 6 Myr ago, which have triggered star formation \citep{FoleyEtal2023}. 
Like the Sco-Cen region, the Orion Nebula contains a superbubble of X-ray emitting, ${}^{26}{\rm Al}$-bearing gas, the Orion-Eridanus superbubble \citep{ReynoldsOgden1979} that also may have triggered star formation across this region of space.

The question of how normal the {Sun's ${}^{26}{\rm Al}$ abundance was} can be answered in this context.
{Contamination by Wolf-Rayet stars resembles the model advocated by \citet{DwarkadasEtal2017}, formation at the edge of a bubble launched by Wolf-Rayet stars, except that in that model the Sun would be rare, corresponding only to material in the swept-up shell.
In our interpretation, like others, the Wolf-Rayet stars contaminate the molecular clouds before star formation, making ${}^{26}{\rm Al}$ more common.}
In their model of how stars form at the periphery of a massive star-forming region,
\citet{GaidosEtal2009} found that the modal abundance would be ${}^{26}{\rm Al} / {}^{27}{\rm Al}$
$\sim 10^{-6} - 10^{-5}$, compared to the initial value in the Solar System, $5 \times 10^{-5}$.
In their whole-galaxy simulations of ${}^{26}{\rm Al}$ enrichment in clouds, 
\citet{FujimotoEtal2018}
found similar values, ${}^{26}{\rm Al} / {}^{27}{\rm Al}$ $\sim 10^{-5}$.
The box model of \citet{Young2020} predicts a modal abundance very similar to the Solar System value, but with a range of values from factor-of-5 lower to a factor-of-5 larger.
In all models, stars like the Sun are predicted to have been born with an initial
${}^{26}{\rm Al}/{}^{27}{\rm Al}$
value similar to, or perhaps an order of magnitude smaller than, the Solar System initial value,
with a range of probable values extending perhaps an order of magnitude both smaller and larger.
Of course, some fraction ($< 20\%$; \cite{LadaLada2003}) of stars form in low-mass star-forming
regions ($N \sim 10^2$ or lower) and may acquire relatively little ${}^{26}{\rm Al}$.

If the Sun formed in a region like the Orion Nebula, late infall might be possible, although it is not {at all} clear this would happen.
Late infall onto the Sun's protoplanetary disk would be a likely outcome if the Sun had formed in a low-mass star-forming region like Taurus-Auriga \citep{PinedaEtal2023}.
It would be very unlikely if it had formed in a high-mass star-forming region like the Carina Nebula; whether the Sun was triggered to form or not, proximity to gas experiencing high $G_0$ implies the Sun's disk would have been shut off from accreting gas within $\sim 10^5$ yr timescales.
For an intermediate-mass region like the Orion Nebula, especially 2 pc from the center of the Orion Nebula Cluster, late infall along streamers has not been observed, but cannot entirely be ruled out.
However, the fact that the Sun apparently formed at the edge of a photodissociation region (PDR) suggests that it quickly (within a few $\times 10^5$ yr) would be uncovered by the ionization front, as in other star-forming regions \citep{SniderEtal2009}.
Even presuming the Sun accreted late material from its molecular cloud, it is not at all clear that it would be compositionally distinct from earlier-accreted material.
In 1 Myr, diffusion due to hydromagnetic turbulence should homogenize material over lengthscales $> 0.3$ pc \citep{PanEtal2012} and is further homogenized during accretion
\citep{KuffmeierEtal2016}.


Based on the evidence presented here, we conclude the Sun formed in a region like the outskirts ($>$ 2 pc from the center) of the Orion Nebula, a region characterized by $G_0 \sim 10^3$, stellar densities $\approx 100 \, M_{\odot} \, {\rm pc}^{-3}$, in a cluster that will remain gravitationally bound for $\approx 100$ Myr.
Such a region would contain $N \sim 10^3 - 10^4$ stars. 
Some estimates of the Sun's birth cluster size are slightly larger (e.g., \cite{PfalznerVincke2020}; see review by
\cite{BerginEtal2023}), mostly because they do not account for the fact that the gas disk could be much larger than the dust disk outer edge at 47 AU (allowing for smaller $G_0$); and especially because they demand proximity to a Wolf-Rayet star or supernova to acquire sufficient ${}^{26}{\rm Al}$, rather than considering previous contamination of the molecular cloud as the source.
Other estimates \citep{ForbesEtal2021} are smaller, because they do not account for the need for the Sun to remain in a bound cluster for 100 Myr. 
Formation in a region like the Orion Nebula is consistent with the Sun's disk size, disk dynamics, Jupiter's abundances, the existence of detached Kuiper belt objects yet the relative non-disturbance of cold classical Kuiper belt objects, the generation of isotopically fractionated reservoirs of oxygen and sulfur in the solar nebula, and the abundances of short-lived radionuclides.
Late ($>$ a few $\times 10^5$ yr) infall cannot be ruled out, but seems unlikely, and should be judged in this context.

\bmhead{Acknowledgments}

The authors are grateful to the organizers of the International Space Science Institute workshop ``Evolution of the Solar System: Constraints from Meteorites (June 5-9, 2023, Bern, Switzerland).

\vspace{0.2in}
\noindent
The authors declare no competing interests.

%
%






\bibliography{SunsBirth1.bib}

\begin{thebibliography}{122}
\providecommand{\natexlab}[1]{#1}
\providecommand{\url}[1]{{#1}}
\providecommand{\urlprefix}{URL }
\providecommand{\doi}[1]{\url{https://doi.org/#1}}
\providecommand{\eprint}[2][]{\url{#2}}
 \bibcommenthead

\bibitem[{{Adams}(2010)}]{Adams2010}
{Adams} FC (2010) {The Birth Environment of the Solar System}. Ann Rev Astron \& Astrophys 48:47--85. \doi{10.1146/annurev-astro-081309-130830}, {\href{https://arxiv.org/abs/1001.5444}{{arXiv:1001.5444}}} {[astro-ph.SR]}

\bibitem[{{Adams} et~al(2004){Adams}, {Hollenbach}, {Laughlin}, and {Gorti}}]{AdamsEtal2004}
{Adams} FC, {Hollenbach} D, {Laughlin} G, et~al (2004) {Photoevaporation of Circumstellar Disks Due to External Far-Ultraviolet Radiation in Stellar Aggregates}. ApJ 611(1):360--379. \doi{10.1086/421989}, {\href{https://arxiv.org/abs/astro-ph/0404383}{{arXiv:astro-ph/0404383}}} {[astro-ph]}

\bibitem[{{Alexander} et~al(2017){Alexander}, {Nittler}, {Davidson}, and {Ciesla}}]{AlexanderEtal2017}
{Alexander} CMO, {Nittler} LR, {Davidson} J, et~al (2017) {Measuring the level of interstellar inheritance in the solar protoplanetary disk}. Meteoritics \& Planet Sci 52(9):1797--1821. \doi{10.1111/maps.12891}

\bibitem[{{Ansdell} et~al(2018){Ansdell}, {Williams}, {Trapman}, {van Terwisga}, {Facchini}, {Manara}, {van der Marel}, {Miotello}, {Tazzari}, {Hogerheijde}, {Guidi}, {Testi}, and {van Dishoeck}}]{AnsdellEtal2018}
{Ansdell} M, {Williams} JP, {Trapman} L, et~al (2018) {ALMA Survey of Lupus Protoplanetary Disks. II. Gas Disk Radii}. ApJ 859(1):21. \doi{10.3847/1538-4357/aab890}, {\href{https://arxiv.org/abs/1803.05923}{{arXiv:1803.05923}}} {[astro-ph.EP]}

\bibitem[{{Arzoumanian} et~al(2023){Arzoumanian}, {Arakawa}, {Kobayashi}, {Iwasaki}, {Fukuda}, {Mori}, {Hirai}, {Kunitomo}, {Kumar}, and {Kokubo}}]{ArzoumanianEtal2023}
{Arzoumanian} D, {Arakawa} S, {Kobayashi} MIN, et~al (2023) {Insights on the Sun Birth Environment in the Context of Star Cluster Formation in Hub-Filament Systems}. ApJLett 947(2):L29. \doi{10.3847/2041-8213/acc849}, {\href{https://arxiv.org/abs/2303.15695}{{arXiv:2303.15695}}} {[astro-ph.GA]}

\bibitem[{{Asplund} et~al(2009){Asplund}, {Grevesse}, {Sauval}, and {Scott}}]{AsplundEtal2009}
{Asplund} M, {Grevesse} N, {Sauval} AJ, et~al (2009) {The Chemical Composition of the Sun}. Ann Rev Astron \& Astrophys 47(1):481--522. \doi{10.1146/annurev.astro.46.060407.145222}, {\href{https://arxiv.org/abs/0909.0948}{{arXiv:0909.0948}}} {[astro-ph.SR]}

\bibitem[{{Bally}(2008)}]{Bally2008}
{Bally} J (2008) {Star formation in Orion and beyond}. In: Star Formation Across the Milky Way Galaxy, p~7

\bibitem[{{Bally} and {Langer}(1982)}]{BallyLanger1982}
{Bally} J, {Langer} WD (1982) {Isotope-selective photodestruction of carbon monoxide.} ApJ 255:143--148. \doi{10.1086/159812}

\bibitem[{{Bar-Nun} et~al(1988){Bar-Nun}, {Kleinfeld}, and {Kochavi}}]{BarNunEtal1988}
{Bar-Nun} A, {Kleinfeld} I, {Kochavi} E (1988) {Trapping of gas mixtures by amorphous water ice}. Phys Rev B 38(11):7749--7754. \doi{10.1103/PhysRevB.38.7749}

\bibitem[{{Barboni} et~al(2017){Barboni}, {Boehnke}, {Keller}, {Kohl}, {Schoene}, {Young}, and {McKeegan}}]{BarboniEtal2017}
{Barboni} M, {Boehnke} P, {Keller} B, et~al (2017) {Early formation of the Moon 4.51 billion years ago}. Science Advances 3(1):e1602365. \doi{10.1126/sciadv.1602365}

\bibitem[{{Batygin} et~al(2020){Batygin}, {Adams}, {Batygin}, and {Petigura}}]{BatyginEtal2020}
{Batygin} K, {Adams} FC, {Batygin} YK, et~al (2020) {Dynamics of Planetary Systems within Star Clusters: Aspects of the Solar System's Early Evolution}. AstronJ 159(3):101. \doi{10.3847/1538-3881/ab665d}, {\href{https://arxiv.org/abs/2002.05656}{{arXiv:2002.05656}}} {[astro-ph.EP]}

\bibitem[{{Bergin} et~al(2023){Bergin}, {Alexander}, {Drozdovskaya}, {Gounelle}, and {Pfalzner}}]{BerginEtal2023}
{Bergin} EA, {Alexander} C, {Drozdovskaya} M, et~al (2023) {Interstellar Heritage and the Birth Environment of the Solar System}. arXiv e-prints arXiv:2301.05212. \doi{10.48550/arXiv.2301.05212}, {\href{https://arxiv.org/abs/2301.05212}{{arXiv:2301.05212}}} {[astro-ph.EP]}

\bibitem[{{Bierson} and {Nimmo}(2019)}]{BiersonNimmo2019}
{Bierson} CJ, {Nimmo} F (2019) {Using the density of Kuiper Belt Objects to constrain their composition and formation history}. Icarus 326:10--17. \doi{10.1016/j.icarus.2019.01.027}

\bibitem[{{Brasser} et~al(2006){Brasser}, {Duncan}, and {Levison}}]{BrasserEtal2006}
{Brasser} R, {Duncan} MJ, {Levison} HF (2006) {Embedded star clusters and the formation of the Oort Cloud}. Icarus 184(1):59--82. \doi{10.1016/j.icarus.2006.04.010}

\bibitem[{{Cameron} and {Truran}(1977)}]{CameronTruran1977}
{Cameron} AGW, {Truran} JW (1977) {The Supernova Trigger for Formation of the Solar System}. Icarus 30(3):447--461. \doi{10.1016/0019-1035(77)90101-4}

\bibitem[{{Chakraborty} et~al(2013){Chakraborty}, {Yanchulova}, and {Thiemens}}]{ChakrabortyEtal2013}
{Chakraborty} S, {Yanchulova} P, {Thiemens} MH (2013) {Mass-Independent Oxygen Isotopic Partitioning During Gas-Phase SiO$_{2}$ Formation}. Science 342(6157):463--466. \doi{10.1126/science.1242237}

\bibitem[{{Ciesla} et~al(2015){Ciesla}, {Mulders}, {Pascucci}, and {Apai}}]{CieslaEtal2015}
{Ciesla} FJ, {Mulders} GD, {Pascucci} I, et~al (2015) {Volatile Delivery to Planets from Water-rich Planetesimals around Low Mass Stars}. ApJ 804(1):9. \doi{10.1088/0004-637X/804/1/9}, {\href{https://arxiv.org/abs/1502.07412}{{arXiv:1502.07412}}} {[astro-ph.EP]}

\bibitem[{{Cleeves} et~al(2016){Cleeves}, {{\"O}berg}, {Wilner}, {Huang}, {Loomis}, {Andrews}, and {Czekala}}]{CleevesEtal2016}
{Cleeves} LI, {{\"O}berg} KI, {Wilner} DJ, et~al (2016) {The Coupled Physical Structure of Gas and Dust in the IM Lup Protoplanetary Disk}. ApJ 832(2):110. \doi{10.3847/0004-637X/832/2/110}, {\href{https://arxiv.org/abs/1610.00715}{{arXiv:1610.00715}}} {[astro-ph.SR]}

\bibitem[{{Da Rio} et~al(2010){Da Rio}, {Robberto}, {Soderblom}, {Panagia}, {Hillenbrand}, {Palla}, and {Stassun}}]{DaRioEtal2010}
{Da Rio} N, {Robberto} M, {Soderblom} DR, et~al (2010) {A Multi-color Optical Survey of the Orion Nebula Cluster. II. The H-R Diagram}. ApJ 722(2):1092--1114. \doi{10.1088/0004-637X/722/2/1092}, {\href{https://arxiv.org/abs/1008.1265}{{arXiv:1008.1265}}} {[astro-ph.GA]}

\bibitem[{{Dale} et~al(2007){Dale}, {Bonnell}, and {Whitworth}}]{DaleEtal2007}
{Dale} JE, {Bonnell} IA, {Whitworth} AP (2007) {Ionization-induced star formation - I. The collect-and-collapse model}. Mon Not Roy Astron Soc 375(4):1291--1298. \doi{10.1111/j.1365-2966.2006.11368.x}, {\href{https://arxiv.org/abs/astro-ph/0612128}{{arXiv:astro-ph/0612128}}} {[astro-ph]}

\bibitem[{{de Sousa} et~al(2020){de Sousa}, {Morbidelli}, {Raymond}, {Izidoro}, {Gomes}, and {Vieira Neto}}]{deSousaEtal2020}
{de Sousa} RR, {Morbidelli} A, {Raymond} SN, et~al (2020) {Dynamical evidence for an early giant planet instability}. Icarus 339:113605. \doi{10.1016/j.icarus.2019.113605}

\bibitem[{{Desch}(2007)}]{Desch2007}
{Desch} SJ (2007) {Mass Distribution and Planet Formation in the Solar Nebula}. ApJ 671(1):878--893. \doi{10.1086/522825}

\bibitem[{{Desch} et~al(2017){Desch}, {Estrada}, {Kalyaan}, and {Cuzzi}}]{DeschEtal2017}
{Desch} SJ, {Estrada} PR, {Kalyaan} A, et~al (2017) {Formulas for Radial Transport in Protoplanetary Disks}. The Astrophysical Journal 840(2):86. \doi{10.3847/1538-4357/aa6bfb}, {\href{https://arxiv.org/abs/1704.01267}{{arXiv:1704.01267}}} {[astro-ph.EP]}

\bibitem[{{Desch} et~al(2018){Desch}, {Kalyaan}, and {O'D. Alexander}}]{DeschEtal2018}
{Desch} SJ, {Kalyaan} A, {O'D. Alexander} CM (2018) {The Effect of Jupiter's Formation on the Distribution of Refractory Elements and Inclusions in Meteorites}. ApJSuppl 238(1):11. \doi{10.3847/1538-4365/aad95f}, {\href{https://arxiv.org/abs/1710.03809}{{arXiv:1710.03809}}} {[astro-ph.EP]}

\bibitem[{{Desch} et~al(2023{\natexlab{a}}){Desch}, {Dunlap}, {Dunham}, {Williams}, and {Mane}}]{DeschEtal2023a}
{Desch} SJ, {Dunlap} DR, {Dunham} ET, et~al (2023{\natexlab{a}}) {Statistical chronometry of meteorites. I. A Test of $^{26}$Al homogeneity and the Pb-Pb age of the solar system's t = 0}. Icarus 402:115607. \doi{10.1016/j.icarus.2023.115607}, {\href{https://arxiv.org/abs/2212.00390}{{arXiv:2212.00390}}} {[astro-ph.EP]}

\bibitem[{{Desch} et~al(2023{\natexlab{b}}){Desch}, {Dunlap}, {Williams}, {Mane}, and {Dunham}}]{DeschEtal2023b}
{Desch} SJ, {Dunlap} DR, {Williams} CD, et~al (2023{\natexlab{b}}) {Statistical chronometry of Meteorites: II. Initial abundances and homogeneity of short-lived radionuclides}. Icarus 402:115611. \doi{10.1016/j.icarus.2023.115611}, {\href{https://arxiv.org/abs/2212.00145}{{arXiv:2212.00145}}} {[astro-ph.EP]}

\bibitem[{{Desch} et~al(2023{\natexlab{c}}){Desch}, {Young}, {Dunham}, {Fujimoto}, and {Dunlap}}]{DeschEtal2023c}
{Desch} SJ, {Young} ED, {Dunham} ET, et~al (2023{\natexlab{c}}) {Short-Lived Radionuclides in Meteorites and the Sun's Birth Environment}. In: {Inutsuka} S, {Aikawa} Y, {Muto} T, et~al (eds) Astronomical Society of the Pacific Conference Series, p 759

\bibitem[{{Diehl}(2002)}]{Diehl2002}
{Diehl} R (2002) {$^{26}$Al production in the Vela and Orion regions}. New Astronomy Reviews 46(8-10):547--552. \doi{10.1016/S1387-6473(02)00199-9}

\bibitem[{{Dunham} et~al(2022){Dunham}, {Wadhwa}, {Desch}, {Liu}, {Fukuda}, {Kita}, {Hertwig}, {Hervig}, {Defouilloy}, {Simon}, {Davidson}, {Schrader}, and {Fujimoto}}]{DunhamEtal2022}
{Dunham} ET, {Wadhwa} M, {Desch} SJ, et~al (2022) {Uniform initial $^{10}$Be/$^{9}$Be inferred from refractory inclusions in CV3, CO3, CR2, and CH/CB chondrites}. GeochimCosmochimActa 324:194--220. \doi{10.1016/j.gca.2022.02.002}

\bibitem[{{Dwarkadas} et~al(2017){Dwarkadas}, {Dauphas}, {Meyer}, {Boyajian}, and {Bojazi}}]{DwarkadasEtal2017}
{Dwarkadas} VV, {Dauphas} N, {Meyer} B, et~al (2017) {Triggered Star Formation inside the Shell of a Wolf-Rayet Bubble as the Origin of the Solar System}. ApJ 851(2):147. \doi{10.3847/1538-4357/aa992e}, {\href{https://arxiv.org/abs/1712.10053}{{arXiv:1712.10053}}} {[astro-ph.SR]}

\bibitem[{{Evans} et~al(2009){Evans}, {Dunham}, {J{\o}rgensen}, {Enoch}, {Mer{\'\i}n}, {van Dishoeck}, {Alcal{\'a}}, {Myers}, {Stapelfeldt}, {Huard}, {Allen}, {Harvey}, {van Kempen}, {Blake}, {Koerner}, {Mundy}, {Padgett}, and {Sargent}}]{EvansEtal2009}
{Evans} NJ, {Dunham} MM, {J{\o}rgensen} JK, et~al (2009) {The Spitzer c2d Legacy Results: Star-Formation Rates and Efficiencies; Evolution and Lifetimes}. ApJSupp 181(2):321--350. \doi{10.1088/0067-0049/181/2/321}, {\href{https://arxiv.org/abs/0811.1059}{{arXiv:0811.1059}}} {[astro-ph]}

\bibitem[{{Flaherty} et~al(2020){Flaherty}, {Hughes}, {Simon}, {Qi}, {Bai}, {Bulatek}, {Andrews}, {Wilner}, and {K{\'o}sp{\'a}l}}]{FlahertyEtal2020}
{Flaherty} K, {Hughes} AM, {Simon} JB, et~al (2020) {Measuring Turbulent Motion in Planet-forming Disks with ALMA: A Detection around DM Tau and Nondetections around MWC 480 and V4046 Sgr}. Ap J 895(2):109. \doi{10.3847/1538-4357/ab8cc5}, {\href{https://arxiv.org/abs/2004.12176}{{arXiv:2004.12176}}} {[astro-ph.SR]}

\bibitem[{{Foley} et~al(2023){Foley}, {Goodman}, {Zucker}, {Forbes}, {Konietzka}, {Swiggum}, {Alves}, {Bally}, {Soler}, {Gro{\ss}schedl}, {Bialy}, {Grudi{\'c}}, {Leike}, and {En{\ss}lin}}]{FoleyEtal2023}
{Foley} MM, {Goodman} A, {Zucker} C, et~al (2023) {A 3D View of Orion. I. Barnard's Loop}. The Astrophysical Journal 947(2):66. \doi{10.3847/1538-4357/acb5f4}, {\href{https://arxiv.org/abs/2212.01405}{{arXiv:2212.01405}}} {[astro-ph.GA]}

\bibitem[{{Forbes} et~al(2021){Forbes}, {Alves}, and {Lin}}]{ForbesEtal2021}
{Forbes} JC, {Alves} J, {Lin} DNC (2021) {A Solar System formation analogue in the Ophiuchus star-forming complex}. Nature Astronomy 5:1009--1016. \doi{10.1038/s41550-021-01442-9}, {\href{https://arxiv.org/abs/2108.09326}{{arXiv:2108.09326}}} {[astro-ph.EP]}

\bibitem[{{Fujimoto} et~al(2018){Fujimoto}, {Krumholz}, and {Tachibana}}]{FujimotoEtal2018}
{Fujimoto} Y, {Krumholz} MR, {Tachibana} S (2018) {Short-lived radioisotopes in meteorites from Galactic-scale correlated star formation}. Mon Not Roy Astron Soc 480(3):4025--4039. \doi{10.1093/mnras/sty2132}, {\href{https://arxiv.org/abs/1802.08695}{{arXiv:1802.08695}}} {[astro-ph.GA]}

\bibitem[{{Fujimoto} et~al(2020){Fujimoto}, {Krumholz}, and {Inutsuka}}]{FujimotoEtal2020}
{Fujimoto} Y, {Krumholz} MR, {Inutsuka} Si (2020) {Distribution and kinematics of $^{26}$Al in the Galactic disc}. Mon Not Roy Astron Soc 497(2):2442--2454. \doi{10.1093/mnras/staa2125}, {\href{https://arxiv.org/abs/2006.03057}{{arXiv:2006.03057}}} {[astro-ph.GA]}

\bibitem[{{Gaidos} et~al(2009){Gaidos}, {Krot}, {Williams}, and {Raymond}}]{GaidosEtal2009}
{Gaidos} E, {Krot} AN, {Williams} JP, et~al (2009) {$^{26}$Al and the Formation of the Solar System from a Molecular Cloud Contaminated by Wolf-Rayet Winds}. ApJ 696(2):1854--1863. \doi{10.1088/0004-637X/696/2/1854}, {\href{https://arxiv.org/abs/0901.3364}{{arXiv:0901.3364}}} {[astro-ph.EP]}

\bibitem[{{Galli} et~al(2020){Galli}, {Bouy}, {Olivares}, {Miret-Roig}, {Vieira}, {Sarro}, {Barrado}, {Berihuete}, {Bertout}, {Bertin}, and {Cuillandre}}]{GalliEtal2020}
{Galli} PAB, {Bouy} H, {Olivares} J, et~al (2020) {Lupus DANCe. Census of stars and 6D structure with Gaia-DR2 data}. Astron \& Astrophys 643:A148. \doi{10.1051/0004-6361/202038717}, {\href{https://arxiv.org/abs/2010.00233}{{arXiv:2010.00233}}} {[astro-ph.SR]}

\bibitem[{{Grimm} and {McSween}(1993)}]{GrimmMcSween1993}
{Grimm} RE, {McSween} HY (1993) {Heliocentric Zoning of the Asteroid Belt by Aluminum-26 Heating}. Science 259(5095):653--655

\bibitem[{{G{\"u}del} et~al(2007){G{\"u}del}, {Briggs}, {Arzner}, {Audard}, {Bouvier}, {Feigelson}, {Franciosini}, {Glauser}, {Grosso}, {Micela}, {Monin}, {Montmerle}, {Padgett}, {Palla}, {Pillitteri}, {Rebull}, {Scelsi}, {Silva}, {Skinner}, {Stelzer}, and {Telleschi}}]{GudelEtal2007}
{G{\"u}del} M, {Briggs} KR, {Arzner} K, et~al (2007) {The XMM-Newton extended survey of the Taurus molecular cloud (XEST)}. Astron \& Astrophys 468(2):353--377. \doi{10.1051/0004-6361:20065724}, {\href{https://arxiv.org/abs/astro-ph/0609160}{{arXiv:astro-ph/0609160}}} {[astro-ph]}

\bibitem[{{Guillot} and {Hueso}(2006)}]{GuillotHueso2006}
{Guillot} T, {Hueso} R (2006) {The composition of Jupiter: sign of a (relatively) late formation in a chemically evolved protosolar disc}. Mon Not Roy Astron Soc 367(1):L47--L51. \doi{10.1111/j.1745-3933.2006.00137.x}, {\href{https://arxiv.org/abs/astro-ph/0601043}{{arXiv:astro-ph/0601043}}} {[astro-ph]}

\bibitem[{{Hacar} et~al(2023){Hacar}, {Clark}, {Heitsch}, {Kainulainen}, {Panopoulou}, {Seifried}, and {Smith}}]{HacarEtal2023}
{Hacar} A, {Clark} SE, {Heitsch} F, et~al (2023) {Initial Conditions for Star Formation: a Physical Description of the Filamentary ISM}. In: {Inutsuka} S, {Aikawa} Y, {Muto} T, et~al (eds) Protostars and Planets VII, p 153, \doi{10.48550/arXiv.2203.09562}, \eprint{2203.09562}

\bibitem[{{Hao} et~al(2023){Hao}, {Xu}, {Hou}, {Lin}, and {Li}}]{HaoEtal2023}
{Hao} CJ, {Xu} Y, {Hou} LG, et~al (2023) {Unveiling the Initial Conditions of Open Star Cluster Formation}. Research in Astronomy and Astrophysics 23(7):075023. \doi{10.1088/1674-4527/acd58d}, {\href{https://arxiv.org/abs/2305.04415}{{arXiv:2305.04415}}} {[astro-ph.GA]}

\bibitem[{{Hartmann} et~al(1998{\natexlab{a}}){Hartmann}, {Calvet}, {Gullbring}, and {D'Alessio}}]{HillenbrandHartmann1998}
{Hartmann} L, {Calvet} N, {Gullbring} E, et~al (1998{\natexlab{a}}) {Accretion and the Evolution of T Tauri Disks}. ApJ 495(1):385--400. \doi{10.1086/305277}

\bibitem[{{Hartmann} et~al(1998{\natexlab{b}}){Hartmann}, {Calvet}, {Gullbring}, and {D'Alessio}}]{HartmannEtal1998}
{Hartmann} L, {Calvet} N, {Gullbring} E, et~al (1998{\natexlab{b}}) {Accretion and the Evolution of T Tauri Disks}. ApJ 495(1):385--400. \doi{10.1086/305277}

\bibitem[{{Hester} and {Desch}(2005)}]{HesterDesch2005}
{Hester} JJ, {Desch} SJ (2005) {Understanding Our Origins: Star Formation in HII Region Environments}. In: {Krot} AN, {Scott} ERD, {Reipurth} B (eds) Chondrites and the Protoplanetary Disk, p 107, \doi{10.48550/arXiv.astro-ph/0506190}, \eprint{astro-ph/0506190}

\bibitem[{{Hollenbach} and {Adams}(2004)}]{HollenbachAdams2004}
{Hollenbach} D, {Adams} FC (2004) {Dispersal of Disks Around Young Stars: Constraints on Kuiper Belt Formation}. In: {Caroff} L, {Moon} LJ, {Backman} D, et~al (eds) Debris Disks and the Formation of Planets, p 168

\bibitem[{{Huss} et~al(2009){Huss}, {Meyer}, {Srinivasan}, {Goswami}, and {Sahijpal}}]{HussEtal2009}
{Huss} GR, {Meyer} BS, {Srinivasan} G, et~al (2009) {Stellar sources of the short-lived radionuclides in the early solar system}. GeochimCosmochimActa 73(17):4922--4945. \doi{10.1016/j.gca.2009.01.039}

\bibitem[{{Ida} et~al(2000){Ida}, {Larwood}, and {Burkert}}]{IdaEtal2000}
{Ida} S, {Larwood} J, {Burkert} A (2000) {Evidence for Early Stellar Encounters in the Orbital Distribution of Edgeworth-Kuiper Belt Objects}. ApJ 528(1):351--356. \doi{10.1086/308179}, {\href{https://arxiv.org/abs/astro-ph/9907217}{{arXiv:astro-ph/9907217}}} {[astro-ph]}

\bibitem[{{Ireland} et~al(2020){Ireland}, {Avila}, {Greenwood}, {Hicks}, and {Bridges}}]{IrelandEtal2020}
{Ireland} TR, {Avila} J, {Greenwood} RC, et~al (2020) {Oxygen Isotopes and Sampling of the Solar System}. Space Sci Rev 216(2):25. \doi{10.1007/s11214-020-0645-3}

\bibitem[{{Jacobsen}(2005)}]{Jacobsen2005}
{Jacobsen} SB (2005) {The Birth of the Solar System in a Molecular Cloud: Evidence from the Isotopic Pattern of Short-lived Nuclides in the Early Solar System}. In: {Krot} AN, {Scott} ERD, {Reipurth} B (eds) Chondrites and the Protoplanetary Disk, p 548

\bibitem[{{Jeffries} et~al(2011){Jeffries}, {Littlefair}, {Naylor}, and {Mayne}}]{JeffriesEtal2011}
{Jeffries} RD, {Littlefair} SP, {Naylor} T, et~al (2011) {No wide spread of stellar ages in the Orion Nebula Cluster}. Mon Not Roy Astron Soc 418(3):1948--1958. \doi{10.1111/j.1365-2966.2011.19613.x}, {\href{https://arxiv.org/abs/1108.2052}{{arXiv:1108.2052}}} {[astro-ph.SR]}

\bibitem[{{Jewitt} et~al(1998){Jewitt}, {Luu}, and {Trujillo}}]{JewittEtal1998}
{Jewitt} D, {Luu} J, {Trujillo} C (1998) {Large Kuiper Belt Objects: The Mauna Kea 8K CCD Survey}. AstronJ 115(5):2125--2135. \doi{10.1086/300335}

\bibitem[{{Jura} et~al(2013){Jura}, {Xu}, and {Young}}]{JuraEtal2013}
{Jura} M, {Xu} S, {Young} ED (2013) {$^{26}$Al in the Early Solar System: Not So Unusual after All}. Ap J Lett 775(2):L41. \doi{10.1088/2041-8205/775/2/L41}, {\href{https://arxiv.org/abs/1308.6325}{{arXiv:1308.6325}}} {[astro-ph.EP]}

\bibitem[{{Kalyaan} et~al(2015){Kalyaan}, {Desch}, and {Monga}}]{KalyaanEtal2015}
{Kalyaan} A, {Desch} SJ, {Monga} N (2015) {External Photoevaporation of the Solar Nebula. II. Effects on Disk Structure and Evolution with Non-uniform Turbulent Viscosity due to the Magnetorotational Instability}. ApJ 815(2):112. \doi{10.1088/0004-637X/815/2/112}, {\href{https://arxiv.org/abs/1511.05620}{{arXiv:1511.05620}}} {[astro-ph.EP]}

\bibitem[{{Karr} and {Martin}(2003)}]{KarrMartin2003}
{Karr} JL, {Martin} PG (2003) {Triggered Star Formation in the W5 H II Region}. ApJ 595(2):900--912. \doi{10.1086/376590}

\bibitem[{{Kenyon} and {Bromley}(2004)}]{KenyonBromley2004}
{Kenyon} SJ, {Bromley} BC (2004) {Stellar encounters as the origin of distant Solar System objects in highly eccentric orbits}. Nature 432(7017):598--602. \doi{10.1038/nature03136}, {\href{https://arxiv.org/abs/astro-ph/0412030}{{arXiv:astro-ph/0412030}}} {[astro-ph]}

\bibitem[{{Krause} et~al(2018){Krause}, {Burkert}, {Diehl}, {Fierlinger}, {Gaczkowski}, {Kroell}, {Ngoumou}, {Roccatagliata}, {Siegert}, and {Preibisch}}]{KrauseEtal2018}
{Krause} MGH, {Burkert} A, {Diehl} R, et~al (2018) {Surround and Squash: the impact of superbubbles on the interstellar medium in Scorpius-Centaurus OB2}. Astron \& Astrophys 619:A120. \doi{10.1051/0004-6361/201732416}, {\href{https://arxiv.org/abs/1808.04788}{{arXiv:1808.04788}}} {[astro-ph.GA]}

\bibitem[{{Krot} et~al(2020){Krot}, {Nagashima}, {Lyons}, {Lee}, and {Bizzarro}}]{KrotEtal2020}
{Krot} AN, {Nagashima} K, {Lyons} JR, et~al (2020) {Oxygen isotopic heterogeneity in the early Solar System inherited from the protosolar molecular cloud}. Science Advances 6(42):eaay2724. \doi{10.1126/sciadv.aay2724}

\bibitem[{{Kroupa}(2001)}]{Kroupa2001}
{Kroupa} P (2001) {On the variation of the initial mass function}. Mon Not Royal Astron Soc 322(2):231--246. \doi{10.1046/j.1365-8711.2001.04022.x}, {\href{https://arxiv.org/abs/astro-ph/0009005}{{arXiv:astro-ph/0009005}}} {[astro-ph]}

\bibitem[{{Kroupa} et~al(2001){Kroupa}, {Aarseth}, and {Hurley}}]{KroupaEtal2001}
{Kroupa} P, {Aarseth} S, {Hurley} J (2001) {The formation of a bound star cluster: from the Orion nebula cluster to the Pleiades}. Mon Not Royal Astron Soc 321(4):699--712. \doi{10.1046/j.1365-8711.2001.04050.x}, {\href{https://arxiv.org/abs/astro-ph/0009470}{{arXiv:astro-ph/0009470}}} {[astro-ph]}

\bibitem[{{Kruijer} et~al(2017){Kruijer}, {Burkhardt}, {Budde}, and {Kleine}}]{KruijerEtal2017}
{Kruijer} TS, {Burkhardt} C, {Budde} G, et~al (2017) {Age of Jupiter inferred from the distinct genetics and formation times of meteorites}. Proceedings of the National Academy of Science 114(26):6712--6716. \doi{10.1073/pnas.1704461114}

\bibitem[{{Kuffmeier} et~al(2016){Kuffmeier}, {Frostholm Mogensen}, {Haugb{\o}lle}, {Bizzarro}, and {Nordlund}}]{KuffmeierEtal2016}
{Kuffmeier} M, {Frostholm Mogensen} T, {Haugb{\o}lle} T, et~al (2016) {Tracking the Distribution of 26Al and 60Fe during the Early Phases of Star and Disk Evolution}. ApJ 826(1):22. \doi{10.3847/0004-637X/826/1/22}, {\href{https://arxiv.org/abs/1605.05008}{{arXiv:1605.05008}}} {[astro-ph.SR]}

\bibitem[{{Lada} and {Lada}(2003)}]{LadaLada2003}
{Lada} CJ, {Lada} EA (2003) {Embedded Clusters in Molecular Clouds}. Ann Rev Astron \& Astrophys 41:57--115. \doi{10.1146/annurev.astro.41.011802.094844}, {\href{https://arxiv.org/abs/astro-ph/0301540}{{arXiv:astro-ph/0301540}}} {[astro-ph]}

\bibitem[{{Lambrechts} and {Johansen}(2014)}]{LambrechtsJohansen2014}
{Lambrechts} M, {Johansen} A (2014) {Forming the cores of giant planets from the radial pebble flux in protoplanetary discs}. Astron \& Astrophys 572:A107. \doi{10.1051/0004-6361/201424343}, {\href{https://arxiv.org/abs/1408.6094}{{arXiv:1408.6094}}} {[astro-ph.EP]}

\bibitem[{{Li} et~al(2020){Li}, {Ingersoll}, {Bolton}, {Levin}, {Janssen}, {Atreya}, {Lunine}, {Steffes}, {Brown}, {Guillot}, {Allison}, {Arballo}, {Bellotti}, {Adumitroaie}, {Gulkis}, {Hodges}, {Li}, {Misra}, {Orton}, {Oyafuso}, {Santos-Costa}, {Waite}, and {Zhang}}]{LiEtal2020}
{Li} C, {Ingersoll} A, {Bolton} S, et~al (2020) {The water abundance in Jupiter's equatorial zone}. Nature Astronomy 4:609--616. \doi{10.1038/s41550-020-1009-3}, {\href{https://arxiv.org/abs/2012.10305}{{arXiv:2012.10305}}} {[astro-ph.EP]}

\bibitem[{{Li} and {Youdin}(2021)}]{LiYoudin2021}
{Li} R, {Youdin} AN (2021) {Thresholds for Particle Clumping by the Streaming Instability}. ApJ 919(2):107. \doi{10.3847/1538-4357/ac0e9f}, {\href{https://arxiv.org/abs/2105.06042}{{arXiv:2105.06042}}} {[astro-ph.EP]}

\bibitem[{{Lichtenberg} et~al(2019){Lichtenberg}, {Golabek}, {Burn}, {Meyer}, {Alibert}, {Gerya}, and {Mordasini}}]{LichtenbergEtal2019}
{Lichtenberg} T, {Golabek} GJ, {Burn} R, et~al (2019) {A water budget dichotomy of rocky protoplanets from $^{26}$Al-heating}. Nature Astronomy 3:307--313. \doi{10.1038/s41550-018-0688-5}, {\href{https://arxiv.org/abs/1902.04026}{{arXiv:1902.04026}}} {[astro-ph.EP]}

\bibitem[{{Liu} et~al(2022){Liu}, {Johansen}, {Lambrechts}, {Bizzarro}, and {Haugb{\o}lle}}]{LiuEtal2022}
{Liu} B, {Johansen} A, {Lambrechts} M, et~al (2022) {Natural separation of two primordial planetary reservoirs in an expanding solar protoplanetary disk}. Science Advances 8(16):eabm3045. \doi{10.1126/sciadv.abm3045}, {\href{https://arxiv.org/abs/2204.10651}{{arXiv:2204.10651}}} {[astro-ph.EP]}

\bibitem[{{Luhman}(2020{\natexlab{a}})}]{Luhman2020a}
{Luhman} KL (2020{\natexlab{a}}) {A Gaia Survey for Young Stars Associated with the Lupus Clouds}. AstronJ 160(4):186. \doi{10.3847/1538-3881/abb12f}, {\href{https://arxiv.org/abs/2009.05123}{{arXiv:2009.05123}}} {[astro-ph.SR]}

\bibitem[{{Luhman}(2020{\natexlab{b}})}]{Luhman2020b}
{Luhman} KL (2020{\natexlab{b}}) {A Gaia Survey for Young Stars Associated with the Lupus Clouds}. AstronJ 160(4):186. \doi{10.3847/1538-3881/abb12f}, {\href{https://arxiv.org/abs/2009.05123}{{arXiv:2009.05123}}} {[astro-ph.SR]}

\bibitem[{{Lyons} and {Young}(2005)}]{LyonsYoung2005}
{Lyons} JR, {Young} ED (2005) {CO self-shielding as the origin of oxygen isotope anomalies in the early solar nebula}. Nature 435(7040):317--320. \doi{10.1038/nature03557}

\bibitem[{{Lyons} et~al(2009){Lyons}, {Bergin}, {Ciesla}, {Davis}, {Desch}, {Hashizume}, and {Lee}}]{LyonsEtal2009}
{Lyons} JR, {Bergin} EA, {Ciesla} FJ, et~al (2009) {Timescales for the evolution of oxygen isotope compositions in the solar nebula}. GeochimCosmochimActa 73(17):4998--5017. \doi{10.1016/j.gca.2009.01.041}

\bibitem[{{Marchington} and {Parker}(2022)}]{MarchingtonParker2022}
{Marchington} B, {Parker} RJ (2022) {The evolution of protoplanetary disc radii and disc masses in star-forming regions}. Mon Not Roy Astron Soc 515(4):5449--5466. \doi{10.1093/mnras/stac2145}, {\href{https://arxiv.org/abs/2208.04330}{{arXiv:2208.04330}}} {[astro-ph.EP]}

\bibitem[{{McCaughrean} and {Pearson}(2023)}]{McCaughreanPearson2023}
{McCaughrean} MJ, {Pearson} SG (2023) {A JWST survey of the Trapezium Cluster \& inner Orion Nebula. I. Observations \& overview}. arXiv e-prints arXiv:2310.03552. \doi{10.48550/arXiv.2310.03552}, {\href{https://arxiv.org/abs/2310.03552}{{arXiv:2310.03552}}} {[astro-ph.SR]}

\bibitem[{{McKeegan} et~al(2011){McKeegan}, {Kallio}, {Heber}, {Jarzebinski}, {Mao}, {Coath}, {Kunihiro}, {Wiens}, {Nordholt}, {Moses}, {Reisenfeld}, {Jurewicz}, and {Burnett}}]{McKeeganEtal2011}
{McKeegan} KD, {Kallio} APA, {Heber} VS, et~al (2011) {The Oxygen Isotopic Composition of the Sun Inferred from Captured Solar Wind}. Science 332(6037):1528. \doi{10.1126/science.1204636}

\bibitem[{{McKinnon} et~al(2020){McKinnon}, {Richardson}, {Marohnic}, {Keane}, {Grundy}, {Hamilton}, {Nesvorn{\'y}}, {Umurhan}, {Lauer}, {Singer}, {Stern}, {Weaver}, {Spencer}, {Buie}, {Moore}, {Kavelaars}, {Lisse}, {Mao}, {Parker}, {Porter}, {Showalter}, {Olkin}, {Cruikshank}, {Elliott}, {Gladstone}, {Parker}, {Verbiscer}, {Young}, and {New Horizons Science Team}}]{McKinnonEtal2020}
{McKinnon} WB, {Richardson} DC, {Marohnic} JC, et~al (2020) {The solar nebula origin of (486958) Arrokoth, a primordial contact binary in the Kuiper Belt}. Science 367(6481):aay6620. \doi{10.1126/science.aay6620}, {\href{https://arxiv.org/abs/2003.05576}{{arXiv:2003.05576}}} {[astro-ph.EP]}

\bibitem[{{Megeath} et~al(2012){Megeath}, {Gutermuth}, {Muzerolle}, {Kryukova}, {Flaherty}, {Hora}, {Allen}, {Hartmann}, {Myers}, {Pipher}, {Stauffer}, {Young}, and {Fazio}}]{MegeathEtal2012}
{Megeath} ST, {Gutermuth} R, {Muzerolle} J, et~al (2012) {The Spitzer Space Telescope Survey of the Orion A and B Molecular Clouds. I. A Census of Dusty Young Stellar Objects and a Study of Their Mid-infrared Variability}. AJ 144(6):192. \doi{10.1088/0004-6256/144/6/192}, {\href{https://arxiv.org/abs/1209.3826}{{arXiv:1209.3826}}} {[astro-ph.GA]}

\bibitem[{{Megeath} et~al(2016){Megeath}, {Gutermuth}, {Muzerolle}, {Kryukova}, {Hora}, {Allen}, {Flaherty}, {Hartmann}, {Myers}, {Pipher}, {Stauffer}, {Young}, and {Fazio}}]{MegeathEtal2016}
{Megeath} ST, {Gutermuth} R, {Muzerolle} J, et~al (2016) {The Spitzer Space Telescope Survey of the Orion A and B Molecular Clouds. II. The Spatial Distribution and Demographics of Dusty Young Stellar Objects}. AJ 151(1):5. \doi{10.3847/0004-6256/151/1/5}, {\href{https://arxiv.org/abs/1511.01202}{{arXiv:1511.01202}}} {[astro-ph.GA]}

\bibitem[{{Mesa-Delgado} et~al(2016){Mesa-Delgado}, {Zapata}, {Henney}, {Puzia}, and {Tsamis}}]{MesaDelgadoEtal2016}
{Mesa-Delgado} A, {Zapata} L, {Henney} WJ, et~al (2016) {Protoplanetary Disks in the Hostile Environment of Carina}. ApJLett 825(1):L16. \doi{10.3847/2041-8205/825/1/L16}, {\href{https://arxiv.org/abs/1605.08809}{{arXiv:1605.08809}}} {[astro-ph.EP]}

\bibitem[{{Miret-Roig} et~al(2022){Miret-Roig}, {Galli}, {Olivares}, {Bouy}, {Alves}, and {Barrado}}]{MiretRoigEtal2022}
{Miret-Roig} N, {Galli} PAB, {Olivares} J, et~al (2022) {The star formation history of Upper Scorpius and Ophiuchus. A 7D picture: positions, kinematics, and dynamical traceback ages}. Astron \& Astrophys 667:A163. \doi{10.1051/0004-6361/202244709}, {\href{https://arxiv.org/abs/2209.12938}{{arXiv:2209.12938}}} {[astro-ph.GA]}

\bibitem[{{Monga} and {Desch}(2015)}]{MongaDesch2015}
{Monga} N, {Desch} S (2015) {External Photoevaporation of the Solar Nebula: Jupiter's Noble Gas Enrichments}. ApJ 798(1):9. \doi{10.1088/0004-637X/798/1/9}, {\href{https://arxiv.org/abs/1410.4870}{{arXiv:1410.4870}}} {[astro-ph.EP]}

\bibitem[{{Morbidelli} and {Levison}(2004)}]{MorbidelliLevison2004}
{Morbidelli} A, {Levison} HF (2004) {Scenarios for the Origin of the Orbits of the Trans-Neptunian Objects 2000 CR$_{105}$ and 2003 VB$_{12}$ (Sedna)}. AstronJ 128(5):2564--2576. \doi{10.1086/424617}, {\href{https://arxiv.org/abs/astro-ph/0403358}{{arXiv:astro-ph/0403358}}} {[astro-ph]}

\bibitem[{{Nesvorn{\'y}} et~al(2023){Nesvorn{\'y}}, {Bernardinelli}, {Vokrouhlick{\'y}}, and {Batygin}}]{NesvornyEtal2023}
{Nesvorn{\'y}} D, {Bernardinelli} P, {Vokrouhlick{\'y}} D, et~al (2023) {Radial distribution of distant trans-Neptunian objects points to Sun's formation in a stellar cluster}. Icarus 406:115738. \doi{10.1016/j.icarus.2023.115738}, {\href{https://arxiv.org/abs/2308.11059}{{arXiv:2308.11059}}} {[astro-ph.EP]}

\bibitem[{{Noviello} et~al(2022){Noviello}, {Desch}, {Neveu}, {Proudfoot}, and {Sonnett}}]{NovielloEtal2022}
{Noviello} JL, {Desch} SJ, {Neveu} M, et~al (2022) {Let It Go: Geophysically Driven Ejection of the Haumea Family Members}. Planet Sci J 3(9):225. \doi{10.3847/PSJ/ac8e03}

\bibitem[{{Ogliore} et~al(2011){Ogliore}, {Huss}, and {Nagashima}}]{OglioreEtal2011}
{Ogliore} RC, {Huss} GR, {Nagashima} K (2011) {Ratio estimation in SIMS analysis}. Nuclear Instruments and Methods in Physics Research B 269(17):1910--1918. \doi{10.1016/j.nimb.2011.04.120}, {\href{https://arxiv.org/abs/1106.0797}{{arXiv:1106.0797}}} {[astro-ph.IM]}

\bibitem[{{Ouellette} et~al(2010){Ouellette}, {Desch}, and {Hester}}]{OuelletteEtal2010}
{Ouellette} N, {Desch} SJ, {Hester} JJ (2010) {Injection of Supernova Dust in Nearby Protoplanetary Disks}. ApJ 711(2):597--612. \doi{10.1088/0004-637X/711/2/597}

\bibitem[{{Pan} et~al(2012){Pan}, {Desch}, {Scannapieco}, and {Timmes}}]{PanEtal2012}
{Pan} L, {Desch} SJ, {Scannapieco} E, et~al (2012) {Mixing of Clumpy Supernova Ejecta into Molecular Clouds}. ApJ 756(1):102. \doi{10.1088/0004-637X/756/1/102}, {\href{https://arxiv.org/abs/1206.6516}{{arXiv:1206.6516}}} {[astro-ph.SR]}

\bibitem[{{Parker} et~al(2012){Parker}, {Maschberger}, and {Alves de Oliveira}}]{ParkerEtal2012}
{Parker} RJ, {Maschberger} T, {Alves de Oliveira} C (2012) {A search for mass segregation of stars and brown dwarfs in {\ensuremath{\rho}} Ophiuchi}. Mon Not Roy Astron Soc 426(4):3079--3085. \doi{10.1111/j.1365-2966.2012.21790.x}, {\href{https://arxiv.org/abs/1208.0005}{{arXiv:1208.0005}}} {[astro-ph.GA]}

\bibitem[{{Parker} et~al(2023){Parker}, {Lichtenberg}, {Patel}, {Polius}, and {Ridsdill-Smith}}]{ParkerEtal2023}
{Parker} RJ, {Lichtenberg} T, {Patel} M, et~al (2023) {Short-lived radioisotope enrichment in star-forming regions from stellar winds and supernovae}. Mon Not Roy Astron Soc 521(4):4838--4851. \doi{10.1093/mnras/stad871}, {\href{https://arxiv.org/abs/2303.11393}{{arXiv:2303.11393}}} {[astro-ph.EP]}

\bibitem[{{Pfalzner} and {Vincke}(2020)}]{PfalznerVincke2020}
{Pfalzner} S, {Vincke} K (2020) {Cradle(s) of the Sun}. The Astrophysical Journal 897(1):60. \doi{10.3847/1538-4357/ab9533}, {\href{https://arxiv.org/abs/2005.11260}{{arXiv:2005.11260}}} {[astro-ph.EP]}

\bibitem[{{Pineda} et~al(2023){Pineda}, {Arzoumanian}, {Andre}, {Friesen}, {Zavagno}, {Clarke}, {Inoue}, {Chen}, {Lee}, {Soler}, and {Kuffmeier}}]{PinedaEtal2023}
{Pineda} JE, {Arzoumanian} D, {Andre} P, et~al (2023) {From Bubbles and Filaments to Cores and Disks: Gas Gathering and Growth of Structure Leading to the Formation of Stellar Systems}. In: {Inutsuka} S, {Aikawa} Y, {Muto} T, et~al (eds) Protostars and Planets VII, p 233, \doi{10.48550/arXiv.2205.03935}, \eprint{2205.03935}

\bibitem[{{Pinte} et~al(2016){Pinte}, {Dent}, {M{\'e}nard}, {Hales}, {Hill}, {Cortes}, and {de Gregorio-Monsalvo}}]{PinteEtal2016}
{Pinte} C, {Dent} WRF, {M{\'e}nard} F, et~al (2016) {Dust and Gas in the Disk of HL Tauri: Surface Density, Dust Settling, and Dust-to-gas Ratio}. Ap J 816(1):25. \doi{10.3847/0004-637X/816/1/25}, {\href{https://arxiv.org/abs/1508.00584}{{arXiv:1508.00584}}} {[astro-ph.SR]}

\bibitem[{{Portegies Zwart}(2019)}]{PortegiesZwart2019}
{Portegies Zwart} S (2019) {The formation of solar-system analogs in young star clusters}. Astron \& Astrophys 622:A69. \doi{10.1051/0004-6361/201833974}, {\href{https://arxiv.org/abs/1810.12934}{{arXiv:1810.12934}}} {[astro-ph.EP]}

\bibitem[{{Ratzenb{\"o}ck} et~al(2023){Ratzenb{\"o}ck}, {Gro{\ss}schedl}, {Alves}, {Miret-Roig}, {Bomze}, {Forbes}, {Goodman}, {Hacar}, {Lin}, {Meingast}, {M{\"o}ller}, {Piecka}, {Posch}, {Rottensteiner}, {Swiggum}, and {Zucker}}]{RatzenbockEtal2023}
{Ratzenb{\"o}ck} S, {Gro{\ss}schedl} JE, {Alves} J, et~al (2023) {The star formation history of the Sco-Cen association. Coherent star formation patterns in space and time}. Astron \& Astrophys 678:A71. \doi{10.1051/0004-6361/202346901}, {\href{https://arxiv.org/abs/2302.07853}{{arXiv:2302.07853}}} {[astro-ph.SR]}

\bibitem[{{Reynolds} and {Ogden}(1979)}]{ReynoldsOgden1979}
{Reynolds} RJ, {Ogden} PM (1979) {Optical evidence for a very large, expanding shell associated with the I Orion OB association, Barnard's loop, and the high galactic latitude Halpha filaments in Eridanus.} The Astrophysical Journal 229:942--953. \doi{10.1086/157028}

\bibitem[{{Sakamoto} et~al(2007){Sakamoto}, {Seto}, {Itoh}, {Kuramoto}, {Fujino}, {Nagashima}, {Krot}, and {Yurimoto}}]{SakamotoEtal2007}
{Sakamoto} N, {Seto} Y, {Itoh} S, et~al (2007) {Remnants of the Early Solar System Water Enriched in Heavy Oxygen Isotopes}. Science 317(5835):231. \doi{10.1126/science.1142021}

\bibitem[{{Sengupta} et~al(2024){Sengupta}, {Cuzzi}, {Umurhan}, and {Lyra}}]{SenguptaEtal2024}
{Sengupta} D, {Cuzzi} JN, {Umurhan} OM, et~al (2024) {Length and Velocity Scales in Protoplanetary Disk Turbulence}. Ap J 966(1):90. \doi{10.3847/1538-4357/ad2c89}, {\href{https://arxiv.org/abs/2402.15475}{{arXiv:2402.15475}}} {[astro-ph.EP]}

\bibitem[{{Smith} et~al(2009){Smith}, {Pontoppidan}, {Young}, {Morris}, and {van Dishoeck}}]{SmithEtal2009}
{Smith} RL, {Pontoppidan} KM, {Young} ED, et~al (2009) {High-Precision C$^{17}$O, C$^{18}$O, and C$^{16}$O Measurements in Young Stellar Objects: Analogues for Co Self-shielding in the Early Solar System}. Ap J 701(1):163--175. \doi{10.1088/0004-637X/701/1/163}, {\href{https://arxiv.org/abs/0906.1024}{{arXiv:0906.1024}}} {[astro-ph.EP]}

\bibitem[{{Snider} et~al(2009){Snider}, {Hester}, {Desch}, {Healy}, and {Bally}}]{SniderEtal2009}
{Snider} KD, {Hester} JJ, {Desch} SJ, et~al (2009) {Spitzer Observations of The H II Region NGC 2467: An Analysis of Triggered Star Formation}. ApJ 700(1):506--522. \doi{10.1088/0004-637X/700/1/506}, {\href{https://arxiv.org/abs/0711.1515}{{arXiv:0711.1515}}} {[astro-ph]}

\bibitem[{{St{\"o}rzer} and {Hollenbach}(1999)}]{StorzerHollenbach1999}
{St{\"o}rzer} H, {Hollenbach} D (1999) {Photodissociation Region Models of Photoevaporating Circumstellar Disks and Application to the Proplyds in Orion}. {\it The Astrophysical Journal} 515(2):669--684. \doi{10.1086/307055}

\bibitem[{{Tachibana} et~al(2006){Tachibana}, {Huss}, {Kita}, {Shimoda}, and {Morishita}}]{TachibanaEtal2006}
{Tachibana} S, {Huss} GR, {Kita} NT, et~al (2006) {$^{60}$Fe in Chondrites: Debris from a Nearby Supernova in the Early Solar System?} Ap J Lett 639(2):L87--L90. \doi{10.1086/503201}

\bibitem[{{Tang} and {Dauphas}(2012)}]{TangDauphas2012}
{Tang} H, {Dauphas} N (2012) {Abundance, distribution, and origin of $^{60}$Fe in the solar protoplanetary disk}. Earth and Planetary Science Letters 359:248--263. \doi{10.1016/j.epsl.2012.10.011}, {\href{https://arxiv.org/abs/1212.1490}{{arXiv:1212.1490}}} {[astro-ph.EP]}

\bibitem[{{Tatischeff} et~al(2014){Tatischeff}, {Duprat}, and {de S{\'e}r{\'e}ville}}]{TatischeffEtal2014}
{Tatischeff} V, {Duprat} J, {de S{\'e}r{\'e}ville} N (2014) {Light-element Nucleosynthesis in a Molecular Cloud Interacting with a Supernova Remnant and the Origin of Beryllium-10 in the Protosolar Nebula}. ApJ 796(2):124. \doi{10.1088/0004-637X/796/2/124}, {\href{https://arxiv.org/abs/1410.1455}{{arXiv:1410.1455}}} {[astro-ph.HE]}

\bibitem[{{Tazzari} et~al(2017){Tazzari}, {Testi}, {Natta}, {Ansdell}, {Carpenter}, {Guidi}, {Hogerheijde}, {Manara}, {Miotello}, {van der Marel}, {van Dishoeck}, and {Williams}}]{TazzariEtal2017}
{Tazzari} M, {Testi} L, {Natta} A, et~al (2017) {Physical properties of dusty protoplanetary disks in Lupus: evidence for viscous evolution?} Astron \& Astrophys 606:A88. \doi{10.1051/0004-6361/201730890}, {\href{https://arxiv.org/abs/1707.01499}{{arXiv:1707.01499}}} {[astro-ph.EP]}

\bibitem[{{Telus} et~al(2016){Telus}, {Huss}, {Ogliore}, {Nagashima}, {Howard}, {Newville}, and {Tomkins}}]{TelusEtal2016}
{Telus} M, {Huss} GR, {Ogliore} RC, et~al (2016) {Mobility of iron and nickel at low temperatures: Implications for $^{60}$Fe-$^{60}$Ni systematics of chondrules from unequilibrated ordinary chondrites}. Geochim Cosmochim Acta 178:87--105. \doi{10.1016/j.gca.2015.11.046}

\bibitem[{{Trapman} et~al(2020){Trapman}, {Rosotti}, {Bosman}, {Hogerheijde}, and {van Dishoeck}}]{TrapmanEtal2020}
{Trapman} L, {Rosotti} G, {Bosman} AD, et~al (2020) {Observed sizes of planet-forming disks trace viscous spreading}. Astron \& Astrophys 640:A5. \doi{10.1051/0004-6361/202037673}, {\href{https://arxiv.org/abs/2005.11330}{{arXiv:2005.11330}}} {[astro-ph.SR]}

\bibitem[{{Trinquier} et~al(2009){Trinquier}, {Elliott}, {Ulfbeck}, {Coath}, {Krot}, and {Bizzarro}}]{TrinquierEtal2009}
{Trinquier} A, {Elliott} T, {Ulfbeck} D, et~al (2009) {Origin of Nucleosynthetic Isotope Heterogeneity in the Solar Protoplanetary Disk}. Science 324(5925):374. \doi{10.1126/science.1168221}

\bibitem[{{Trujillo} and {Brown}(2001)}]{TrujilloBrown2001}
{Trujillo} CA, {Brown} ME (2001) {The Radial Distribution of the Kuiper Belt}. ApJLett 554(1):L95--L98. \doi{10.1086/320917}

\bibitem[{{Tsiganis} et~al(2005){Tsiganis}, {Gomes}, {Morbidelli}, and {Levison}}]{TsiganisEtal2005}
{Tsiganis} K, {Gomes} R, {Morbidelli} A, et~al (2005) {Origin of the orbital architecture of the giant planets of the Solar System}. Nature 435(7041):459--461. \doi{10.1038/nature03539}

\bibitem[{{Vacher} et~al(2021){Vacher}, {Ogliore}, {Jones}, {Liu}, and {Fike}}]{VacherEtal2021}
{Vacher} LG, {Ogliore} RC, {Jones} C, et~al (2021) {Cosmic symplectite recorded irradiation by nearby massive stars in the solar system's parent molecular cloud}. GeochimCosmochimActa 309:135--150. \doi{10.1016/j.gca.2021.06.026}

\bibitem[{{Wall} et~al(2020){Wall}, {Mac Low}, {McMillan}, {Klessen}, {Portegies Zwart}, and {Pellegrino}}]{WallEtal2020}
{Wall} JE, {Mac Low} MM, {McMillan} SLW, et~al (2020) {Modeling of the Effects of Stellar Feedback during Star Cluster Formation Using a Hybrid Gas and N-Body Method}. ApJ 904(2):192. \doi{10.3847/1538-4357/abc011}, {\href{https://arxiv.org/abs/2003.09011}{{arXiv:2003.09011}}} {[astro-ph.SR]}

\bibitem[{{Warren}(2011)}]{Warren2011}
{Warren} PH (2011) {Stable isotopes and the noncarbonaceous derivation of ureilites, in common with nearly all differentiated planetary materials}. GeochimCosmochimActa 75(22):6912--6926. \doi{10.1016/j.gca.2011.09.011}

\bibitem[{{Williams} and {Gaidos}(2007)}]{WilliamsGaidos2007}
{Williams} JP, {Gaidos} E (2007) {On the Likelihood of Supernova Enrichment of Protoplanetary Disks}. ApJLett 663(1):L33--L36. \doi{10.1086/519972}, {\href{https://arxiv.org/abs/0705.3459}{{arXiv:0705.3459}}} {[astro-ph]}

\bibitem[{{Wilson} and {Militzer}(2010)}]{WilsonMilitzer2010}
{Wilson} HF, {Militzer} B (2010) {Sequestration of Noble Gases in Giant Planet Interiors}. Phys Rev Lett 104(12):121101. \doi{10.1103/PhysRevLett.104.121101}, {\href{https://arxiv.org/abs/1003.5940}{{arXiv:1003.5940}}} {[astro-ph.EP]}

\bibitem[{{Winter} et~al(2018){Winter}, {Clarke}, {Rosotti}, {Ih}, {Facchini}, and {Haworth}}]{WinterEtal2018}
{Winter} AJ, {Clarke} CJ, {Rosotti} G, et~al (2018) {Protoplanetary disc truncation mechanisms in stellar clusters: comparing external photoevaporation and tidal encounters}. Mon Not Roy Astron Soc 478(2):2700--2722. \doi{10.1093/mnras/sty984}, {\href{https://arxiv.org/abs/1804.00013}{{arXiv:1804.00013}}} {[astro-ph.SR]}

\bibitem[{{Youdin} and {Lithwick}(2007)}]{YoudinLithwick2007}
{Youdin} AN, {Lithwick} Y (2007) {Particle stirring in turbulent gas disks: Including orbital oscillations}. Icarus 192(2):588--604. \doi{10.1016/j.icarus.2007.07.012}, {\href{https://arxiv.org/abs/0707.2975}{{arXiv:0707.2975}}} {[astro-ph]}

\bibitem[{{Young}(2007)}]{Young2007}
{Young} ED (2007) {Time-dependent oxygen isotopic effects of CO self shielding across the solar protoplanetary disk}. Earth and Planetary Science Letters 262(3-4):468--483. \doi{10.1016/j.epsl.2007.08.011}

\bibitem[{{Young}(2014)}]{Young2014}
{Young} ED (2014) {Inheritance of solar short- and long-lived radionuclides from molecular clouds and the unexceptional nature of the solar system}. Earth and Planetary Science Letters 392:16--27. \doi{10.1016/j.epsl.2014.02.014}, {\href{https://arxiv.org/abs/1403.0832}{{arXiv:1403.0832}}} {[astro-ph.EP]}

\bibitem[{{Young}(2020)}]{Young2020}
{Young} ED (2020) {The birth environment of the solar system constrained by the relative abundances of the solar radionuclides}. In: {Elmegreen} BG, {T{\'o}th} LV, {G{\"u}del} M (eds) Origins: From the Protosun to the First Steps of Life, pp 70--77, \doi{10.1017/S1743921319001777}, \eprint{1909.06361}

\bibitem[{{Yurimoto} and {Kuramoto}(2004)}]{YurimotoKuramoto2004}
{Yurimoto} H, {Kuramoto} K (2004) {Molecular Cloud Origin for the Oxygen Isotope Heterogeneity in the Solar System}. Science 305(5691):1763--1766. \doi{10.1126/science.1100989}

\bibitem[{{Zolensky} et~al(2006){Zolensky}, {Zega}, {Yano}, {Wirick}, {Westphal}, {Weisberg}, {Weber}, {Warren}, {Velbel}, {Tsuchiyama}, {Tsou}, {Toppani}, {Tomioka}, {Tomeoka}, {Teslich}, {Taheri}, {Susini}, {Stroud}, {Stephan}, {Stadermann}, {Snead}, {Simon}, {Simionovici}, {See}, {Robert}, {Rietmeijer}, {Rao}, {Perronnet}, {Papanastassiou}, {Okudaira}, {Ohsumi}, {Ohnishi}, {Nakamura-Messenger}, {Nakamura}, {Mostefaoui}, {Mikouchi}, {Meibom}, {Matrajt}, {Marcus}, {Leroux}, {Lemelle}, {Le}, {Lanzirotti}, {Langenhorst}, {Krot}, {Keller}, {Kearsley}, {Joswiak}, {Jacob}, {Ishii}, {Harvey}, {Hagiya}, {Grossman}, {Grossman}, {Graham}, {Gounelle}, {Gillet}, {Genge}, {Flynn}, {Ferroir}, {Fallon}, {Ebel}, {Dai}, {Cordier}, {Clark}, {Chi}, {Butterworth}, {Brownlee}, {Bridges}, {Brennan}, {Brearley}, {Bradley}, {Bleuet}, {Bland}, and {Bastien}}]{ZolenskyEtal2006}
{Zolensky} ME, {Zega} TJ, {Yano} H, et~al (2006) {Mineralogy and Petrology of Comet 81P/Wild 2 Nucleus Samples}. Science 314(5806):1735. \doi{10.1126/science.1135842}

\end{thebibliography}

\end{document}